\documentclass[amsmath,amssymb,preprintnumbers,nofootinbib,a4paper,11pt]{article}
\pdfoutput=1

\usepackage[pdftex]{graphicx}
\usepackage{epstopdf}
\usepackage{jheppub}
\usepackage{multirow}               
\usepackage{amsmath,amsthm,latexsym,amssymb,amsfonts,epsfig}
\usepackage{fontenc,layout}
\usepackage{slashed}
\usepackage{color}
\usepackage{subfigure}

\usepackage{hyperref}
\hypersetup{
  colorlinks=true,
  linkcolor=blue,
  urlcolor=blue,
  citecolor=blue 
}

\newcommand{\bea}{\begin{align}}
\newcommand{\eea}{\end{align}}
\newcommand{\beq}{\begin{equation}}
\newcommand{\eeq}{\end{equation}}
\newcommand{\nbea}{\begin{align*}}
\newcommand{\neea}{\end{align*}}
\newcommand{\nbeq}{\begin{equation*}}
\newcommand{\neeq}{\end{equation*}}
\newcommand{\bear}{\begin{eqnarray}}  
\newcommand{\eear}{\end{eqnarray}}  

 \definecolor{cadmiumgreen}{rgb}{0.0, 0.42, 0.24}

\numberwithin{equation}{section}

\title{Light quark Yukawas in triboson final states}

\preprint{\begin{flushright}
CERN-TH-2020-191 \\ IFT-UAM/CSIC-20-158 \\ FTUAM-20-25 \\
KEK-TH-2276
\end{flushright}}

\author[1]{Adam Falkowski,}
\author[2]{Sanmay Ganguly,}
\author[3]{Phillippe Gras,}
\author[4,5]{Jose~Miguel~No,}
\author[6,7]{Kohsaku Tobioka,}
\author[8]{Natascia Vignaroli,}
\author[9,10]{Tevong You}

\affiliation[1]{Universit\'{e} Paris-Saclay, CNRS/IN2P3, IJCLab, 91405 Orsay, France}
\affiliation[2]{Department of Particle Physics \& Astrophysics, Weizmann Institute of Science, Israel}
\affiliation[3]{CEA  Institut  de  Recherche  sur  les  lois  Fondamentales  de  l'Univers,  Universit\'e  Paris-Saclay, Gif-sur-Yvette, France}
\affiliation[4]{Departamento de F\'isica Te\'orica, Universidad Aut\'onoma de Madrid, 28049, Madrid, Spain}
\affiliation[5]{Instituto de F\'isica Te\'orica, IFT-UAM/CSIC,
Cantoblanco, 28049, Madrid, Spain}
\affiliation[6]{Department of Physics, Florida State University, Tallahassee, FL 32306, USA}
\affiliation[7]{High Energy Accelerator Research Organization (KEK), Tsukuba 305-0801, Japan}
\affiliation[8]{Dipartimento di Fisica "E. Fermi", Universit\'a di Pisa, and INFN Pisa, Largo Bruno Pontecorvo 3, 56127 Pisa, Italy}
\affiliation[9]{CERN, Theoretical Physics Department, Geneva, Switzerland}
\affiliation[10]{DAMTP, University of Cambridge, Wilberforce Road, Cambridge, CB3 0WA, UK; Cavendish Laboratory, University of Cambridge, J.J. Thomson Avenue, Cambridge, CB3 0HE, UK}

\abstract{
Triple heavy vector boson production, $p p \to VVV$ $(V = W, Z)$, has recently been observed for the first time.  
We propose that precision measurements of this process provide an excellent probe of the first generation light quark Yukawa couplings. Modified quark interactions with the off-shell Higgs in this process lead to a rapid growth of the partonic cross sections with energy, which manifests in an enhanced $p_T$ distribution of the final state leptons and quarks. We quantify this effect and estimate the  present and future 2$\sigma$ sensitivity to the up, down, and strange Yukawas. 
In particular, we find that HL-LHC can reach $\mathcal{O}(400)$ sensitivity to the down Yukawa relative to the Standard Model value, improving the current sensitivity in this process by a factor of $10$, and which can be further improved to $\mathcal{O}(30)$ at FCC-hh. This is competitive with and complementary to constraints from global fits and other on-shell probes of the first generation Yukawas.
The triboson sensitivity at HL-LHC corresponds to  probing dimension-6 SMEFT operators suppressed by an $\mathcal{O}(1)$ TeV scale, similarly to other LHC Higgs probes.  
}

\begin{document}

\maketitle


\section{Introduction} 

The Higgs mechanism plays a central role in the Standard Model (SM). It leads to spontaneous breaking of the electroweak (EW) symmetry, giving masses to the $W$ and $Z$ bosons, and at the same time it generates masses for the SM quarks and leptons.
One important prediction is the existence of a scalar particle---the Higgs boson---whose Yukawa interactions with each SM fermion have their strength proportional to the fermion's mass. 
Precision studies of the Higgs boson at the LHC are providing spectacular confirmation of this prediction. 
We currently have firm evidence~\cite{ATLAS:2020qdt,CMS:2020gsy} that the Higgs couples to the $3^{\rm rd}$ generation fermions (top, bottom, tau) with the strength predicted by the SM, within ${\cal O}(10\%)$ accuracy. 
There is also preliminary evidence for the Higgs decays to muons~\cite{Aad:2020xfq,Sirunyan:2020two}, 
which is the first experimental verification of the SM Higgs mechanism at the level of the $2^{\rm nd}$ generation fermions. 

Higgs decays to $1^{\rm st}$ and $2^{\rm nd}$ generation quarks are much more difficult to determine experimentally. 
If the SM predictions concerning their magnitude are borne out in nature, it will be very challenging to pinpoint their signatures in current or future hadron colliders. 
However, the SM predictions should not be taken for granted.
The pattern of observed fermion masses is mysterious, which prompts many theorists to suspect a deeper explanation involving new physics beyond the SM.  
That new physics might dramatically alter the Higgs boson couplings to the light quarks. In the absence of deeper theoretical understanding, it is essential to continue sharpening the experimental picture with the help of novel analysis strategies and techniques.

The literature already contains several theoretical ideas to better aim at the Higgs coupling to the charm~\cite{Bodwin:2013gca,Perez:2015aoa,Perez:2015lra, Brivio:2015fxa,Mao:2019hgg,Coyle:2019hvs} and/or lighter~\cite{Kagan:2014ila,Koenig:2015pha,Perez:2015lra,Bishara:2016jga,Soreq:2016rae,Bonner:2016sdg,Yu:2016rvv,Cohen:2017rsk,Alasfar:2019pmn,Aguilar-Saavedra:2020rgo} quarks, and several targeted analyses by the LHC collaborations have appeared~\cite{Aad:2015sda,Aaboud:2016rug,LHCb:2016yxg,Aaboud:2017xnb,Aaboud:2018fhh,Sirunyan:2020mds}.
These have so far relied on processes with on-shell Higgs boson(s).
We propose an alternative probe, where the Higgs boson appears only as an intermediate off-shell particle in triboson production processes. 
Our approach is in line with the programme of {\it measuring Higgs couplings without Higgs bosons} laid out in Ref.~\cite{Henning:2018kys}.
Indeed, the SM multi-boson processes and Higgs physics are intricately intertwined by the delicate cancellations necessary to avoid violation of perturbative unitarity in the high-energy behavior of scattering amplitudes. 
Beyond the SM (BSM), when that cancellation is disrupted, 
the  multi-boson cross sections may rapidly grow with the center-of-mass energy of the collision, thus amplifying the effect of the non-SM perturbation.  
Experimental searches for such energy-growing effects allow one to identify or constrain possible deviations of Higgs couplings from their SM values. 
In particular, they provide a novel handle on the Higgs Yukawa couplings to light quarks---up, down, and strange---as we argue in this paper.

\begin{figure}[t]
\centering
\includegraphics[width=0.26\textwidth]{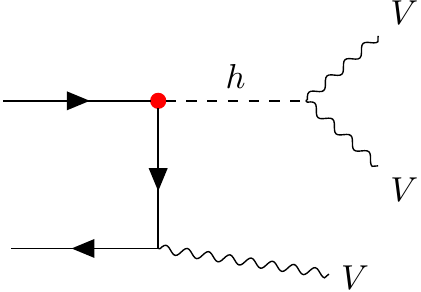} 
\hspace{6mm}
\includegraphics[width=0.26\textwidth]{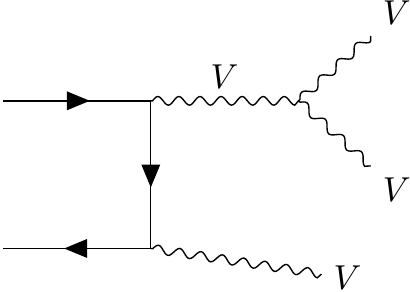} 
\hspace{6mm}
\includegraphics[width=0.26\textwidth]{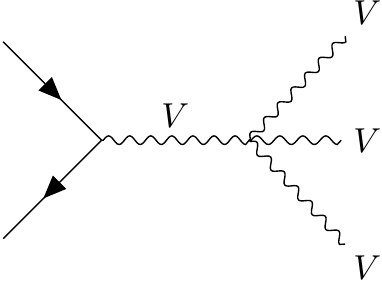} 
\caption{ \small 
Illustrative tree-level Feynman diagrams in the unitary gauge contributing to the triple electroweak gauge boson production, $f \bar f \to VVV$. In the SM, the Higgs exchange diagrams (left) cancel the bad high-energy behavior of the remaining diagrams to ensure that $\sigma(f \bar f \to VVV)$ does not grow with the center-of-mass energy. \label{fig:vvv_diagrams}
} 
\end{figure}

Specifically, we study here the triple EW gauge boson production in hadron colliders,  
$pp \to V V V$, where $V$ stands for an on-shell $W$ or $Z$ gauge boson. 
This process was recently observed for the first time by the CMS collaboration~\cite{CMS:2020gvq,Sirunyan:2020cjp}. 
At the partonic level it is dominated by $q \bar q \to V V V$, which receives a contribution from the diagram with an intermediate Higgs, see Fig.~\ref{fig:vvv_diagrams}.
That diagram plays a crucial role in controlling the high-energy behavior of the  amplitude, 
and any deviations of the $h \bar q q$ (and $h V V$) couplings from the SM value lead to the $q \bar q \to V V V$ cross-section growing quadratically with energy. 
In this work we show that this effect allows one to obtain competitive constraints on the Higgs Yukawa couplings to  the up, down and strange quarks.\footnote{%
The same process also probes the charm and bottom Yukawa couplings, but in this case the effect is suppressed by the small $c$ and $b$ parton distribution functions (PDFs) inside the proton.  
Therefore the resulting constraints are inferior to those obtained from other more direct probes, and we do not discuss them in this paper.  
}

In the next Section we introduce our theoretical framework and notation for modified Yukawa couplings. In Section~\ref{sec:triple} we study the 2-, 3-, and 4-lepton final states and backgrounds of triple EW gauge boson production and estimate their sensitivity to modified up, down and strange Yukawa couplings. 
The comparison with other constraints on the same couplings is discussed in Section~\ref{sec:comparison}, before concluding in Section~\ref{sec:conclusion}.

\section{Theoretical framework} 
\label{sec:framework}

We work in the framework of the SM effective field theory (SMEFT), in which the SM Lagrangian is supplemented with a set of gauge-invariant higher-dimensional operators built from the SM fields. 
The latter encode possible effects of heavy BSM particles on the phenomenology at the EW scale.    
In this work we will focus on a specific subset of dimension-6 operators: 
\begin{equation}
\label{eq:TH_d6}
 {\cal L}_{\rm SMEFT} \supset   \frac{Y_u |H|^2 }{v^2} \bar u_R Q_{1,L} H  
 + \frac{Y_d |H|^2 }{v^2}  \bar d_R H^\dagger Q_{1,L}
  + \frac{Y_s |H|^2 }{v^2}  \bar s_R H^\dagger Q_{2,L}
+{\rm h.c.} \,  ,  
\end{equation}
where $Q_{1,L} = (u_L,d_L)$ and $Q_{2,L} = (c_L,s_L)$ are the left-handed $1^{\rm st}$ and $2^{\rm nd}$ generation SM quark doublets, 
$H$ is the Higgs doublet, and $v \approx 246.22$~GeV is its VEV. 
We assume that the parameters  $Y_q$ are real, 
for simplicity. 
The operators in \eqref{eq:TH_d6} are generated e.g. by integrating out heavy vector-like quarks with masses of order $v/\sqrt{|Y_q|}$ and mixing with the light SM quarks after EW symmetry breaking.  
In restricting to this set of operators, we hereby assume that these
give the dominant non-SM contribution to Higgs phenomenology, 
and neglect possible effects of other dimension-6 operators. 
The Higgs doublet can be parametrized  as 
\begin{equation}
\label{eq:TH_Hdoublet}
H = \frac{1}{\sqrt 2} \begin{pmatrix} i \sqrt 2 G_+ 
\\ v + h + i G_z  \end{pmatrix},    
\end{equation}
where $h$ is the Higgs boson field, and $G_i$ are unphysical Goldstone bosons eaten by the $W$ and $Z$ gauge bosons. By virtue of the equivalence theorem~\cite{Lee:1977eg},
the Goldstone bosons can be in a certain sense identified with the longitudinal polarisations of the $W$ and $Z$ bosons for processes with a characteristic energy scale $E \gg m_Z$. 
This point will prove key to understand the phenomena discussed in this work.

The SMEFT provides a very convenient framework for understanding correlations between various new physics effects.   
In particular, each of the higher-dimensional operators in \eqref{eq:TH_d6} simultaneously contributes  to several observables, such as the Higgs signal strength, double Higgs production, triple EW gauge boson production, etc.
In the following of this section we will show how the modified $h\bar q q$ couplings and the high-energy behavior of the $p p \to VVV$ cross section are correlated.\footnote{%
We note in passing that the same operators also lead to an energy-growing behaviour of the $p p \to W W q j$ cross section. 
The sensitivity of that process to the charm Yukawa was previously studied in Ref.~\cite{Brooijmans:2020yij}, and to the top Yukawa coupling in Refs.~\cite{Henning:2018kys,Maltoni:2019pau,Brooijmans:2020yij}. See also Ref.~\cite{Degrande:2018fog} for a related process with the linear energy growth.}

\begin{table}[h!]
\centering
\begin{tabular}{|c|c|c|c|c|c|c|}
\hline 
& u & d & s & c & b  \\ \hline \hline
$m_q (m_h)$~GeV & 0.0013(1) & 0.0027(1) & 0.0524(4) 
& 0.616(4) & 2.804(8)    \\ \hline  
$y_q^{\rm SM}$ & $7.5 \times 10^{-6}$ & $1.5 \times 10^{-5}$ 
& $3.0 \times 10^{-4}$ & $3.5 \times 10^{-3}$ &  $1.6 \times 10^{-2}$ 
\\ \hline  
${\rm Br }(h \to q \bar q )_{\rm SM}$ & $1.3 \times 10^{-7}$ & $ 5.5 \times 10^{-7}$ &  $2.1 \times 10^{-4}$  &  
 $ 2.9 \times 10^{-2}$  &   0.58
\\ \hline  
\end{tabular}
\caption{\label{tab:ysm}
Quark masses and SM Yukawa coupling values at the Higgs mass scale.  
These have been obtained using the input values at $\mu =2$~GeV: $m_u = 2.3(1)$~MeV,  
$m_d = 4.7(1)$~MeV, $m_s = 92.9(7)$~MeV~\cite{Tanabashi:2018oca}, 
as well as 
$m_c(3~{\rm GeV}) = 0.988(7)$~GeV,
$m_b(m_b) = 4.198(12)$~GeV~\cite{Aoki:2019cca}. 
These have then been evolved up to $\mu = m_h$ using the 4-loop QCD running equations~\cite{Chetyrkin:1997dh,Vermaseren:1997fq} together with the 5-loop running $\alpha_s$~\cite{Baikov:2016tgj} and $\alpha_s(m_Z)=0.1179(10)$~\cite{Tanabashi:2018oca}. 
}
\end{table}

Let us first discuss the relation between the operators in Eq.~\eqref{eq:TH_d6} and the coupling strength between the Higgs boson and  the light quarks.
We parametrize these Yukawa couplings as 
\begin{equation}
\label{eq:TH_yukawa}
{\cal L} \supset  - \frac{h}{v} \sum_{q=u,d,s} m_q \big ( 1 + \delta y_q \big )  \bar q q .
\end{equation}
The relative shifts of the Yukawa couplings with respect to the SM values are encoded in the parameters $\delta y_q$. They are related to the parameters in Eq.~\eqref{eq:TH_d6} by 
\beq
\label{eq:deltayq_def}
\delta y_q  =  - \frac{Y_q}{y_q^{\rm SM}}\,, 
\eeq
with the SM Yukawa coupling defined as $y_q^{\rm SM} \equiv  \sqrt 2  m_q /v$.    
In all of the following we will use $y_q^{\rm SM}$ evaluated at the Higgs mass scale\footnote{We note that, since we will be interested in off-shell Higgs processes, the relevant energy scale $\mu$ at which $y_q^{\rm SM}$ should be evaluated is typically larger than $m_h$, but the renormalization group evolution from $m_h$ to the actual scale probed by the differential distribution generally has a very small impact relative to current uncertainties~\cite{Englert:2014cva}.}, that is to say, calculated using the quark masses evolved up to the renormalization scale $\mu = m_h$.    
The quantity $m_q(m_h)$ differs by an ${\cal O}(2)$ factor from the low-energy value of the corresponding quark mass.  
The numerical values of $m_q(m_h)$ and  $y_q^{\rm SM}$ for light quarks are summarized in Table~\ref{tab:ysm}.

\begin{figure}[h!]
\centering
\includegraphics[width=0.23\textwidth]{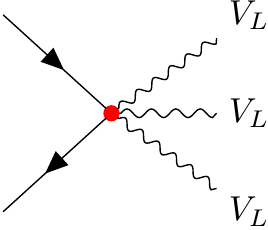} 
\caption{ \small 
The Feynman diagram in the non-unitary gauge that gives the dominant contribution to the ${\cal M}(q \bar q \to G G G)$ amplitude at large center-of-mass energies $\sqrt{s}$.  
By the equivalence theorem, this amplitude is approximately equal to the amplitude for production of longitudinally polarized EW gauge bosons, ${\cal M}(q \bar q \to V_L V_L V_L)$ in the energy range $\sqrt{s} \gg m_Z$. 
\label{fig:vvv_goldstonediagram}
} 
\end{figure}

We move to discussing the connection between the operators in  Eq.~\eqref{eq:TH_d6} and the triple EW gauge boson production. 
As we discussed in Introduction, as soon as $\delta y_q \neq 0$,  the Higgs exchange diagram in Fig.~\ref{fig:vvv_diagrams} no longer regulates the bad high-energy behavior of the remaining diagrams, leading to the quadratic energy growth of the $q \bar q \to VVV$ cross section. 
There is a more transparent way to see it starting from \eqref{eq:TH_d6} and taking advantage of the equivalence theorem. 
Inserting the $H$ parametrization of  Eq.~\eqref{eq:TH_Hdoublet} into  Eq.~\eqref{eq:TH_d6} leads to the contact interactions between two quarks and three Goldstone bosons: 
\begin{eqnarray}
\label{eq:TH_contact}
{\cal L} &  \supset & 
\frac{1}{v^2} \bigg (  G_+ G_- + \frac{1}{2} G_z^2  \bigg ) 
\bigg \{ 
i y_u^{\rm SM }  \delta y_u 
 \left ( \sum_{q' = d,s} \bar u_R  q_L' G_+   - \bar u_R u_L   \frac{G_z}{\sqrt 2} \right )   
\nonumber \\   & +  & 
 i  \sum_{q' = d,s}  y_{q'}^{\rm SM}  \delta y_{q'} 
 \left ( \bar q_R' u_L  G_-  +   \bar q_R'  q_L' \frac{G_z}{\sqrt 2}    \right )   
  + {\rm h.c.} 
  \bigg \} .
 \end{eqnarray}
 These interactions are relevant for the ${\cal M}(q \bar q \to G G G)$ amplitude which, by the equivalence theorem, approximates the high-energy behavior of the  ${\cal M}(q \bar q \to V V V)$ amplitude. 
By dimensional analysis, they contribute as 
${\cal M}(q \bar q \to G G G) \sim {\cal O}(\delta y_q E/v^2)$ for $E \gg m_q$, 
where $E$ is the center-of-mass energy of the process. 
Other tree-level diagrams affecting the same process are more suppressed at high energies because they contain internal propagators.  
Thus, whenever $\delta y_q \neq 0$, 
the diagram in Fig.~\ref{fig:vvv_goldstonediagram} represents the dominant contribution to  ${\cal M}(q \bar q \to G G G)$ at high energies.  
Note that, in this picture, the high-energy behavior is controlled by a {\it single} diagram. 
This is a qualitative simplification compared to the calculation in the unitary gauge, where this high-energy behavior depends on an interplay between several diagrams in  Fig.~\ref{fig:vvv_diagrams}. 
Of course, both calculations are guaranteed to yield the same result, as a consequence of the gauge invariance of the SMEFT.

Treating the quarks and Goldstone bosons as massless, the cross sections for the $q \bar q \to G G G$ processes mediated by the interactions in Eq.~\eqref{eq:TH_contact} are given by the simple analytic expressions  
\begin{eqnarray}
\label{eq:TXS_qqbGzGpGm}
\sigma (q \bar q \to G_z G_+ G_-)   &= &     ( y_q^{\rm SM }  \delta y_q)^2  I(\hat s), 
\qquad I(\hat s) \equiv  \frac{\hat  s}{6144 \pi^3  v^4}  . 
\nonumber \\
\sigma (q \bar q \to 3 G_z)   & =  &  
\frac{3}{2}     ( y_q^{\rm SM }  \delta y_q)^2  I(\hat s)  , 
\nonumber \\
\sigma  (u \bar q' \to G_+  G_z G_z)  +  \sigma  (q' \bar u \to G_-  G_z G_z)   & =  &   
\frac{1}{2}     \left [  ( y_u^{\rm SM }  \delta y_u)^2 +  ( y_{q'}^{\rm SM }  \delta y_{q'})^2 \right ]   I( \hat s), 
\nonumber \\
\sigma  (u \bar q' \to G_+  G_+ G_-) + \sigma  (q' \bar u \to G_-  G_- G_+) & =  &   
 2  \left [  ( y_u^{\rm SM }  \delta y_u)^2 +  ( y_{q'}^{\rm SM }  \delta y_{q'})^2 \right ]   I(\hat s), 
\end{eqnarray} 
with $q = u,d,s$ and $q' = d,s$, and $\sqrt{\hat s}$ is the centre-of-mass energy of the parton level 
quark-antiquark collision.
By the equivalence theorem, 
these expressions approximate the parton-level triple EW gauge boson cross sections
 for $\sqrt{\hat s} \gg m_Z$, 
 with the identification $G_z \to Z_L$ and $G_\pm \to W^\pm_L$. 
 
We can readily make a number of observations that will be important for the subsequent collider analysis:
\vspace{2mm}

\hspace{-5mm} {\it (i)} As expected from the above general arguments, for $\delta y_q \neq 0$ the cross section in all channels grows quadratically with the centre-of-mass energy of the partonic collision. On the other hand, in the SM the triple EW gauge boson cross section is instead suppressed at high energies. Therefore, selecting $VVV$ events with large $\sqrt{\hat s}$ will enhance the sensitivity to $\delta y_q \neq 0$.

\vspace{2mm}
\hspace{-5mm} {\it (ii)} The dimension-6 operators in Eq.~\eqref{eq:TH_d6} lead to a characteristic signal pattern in different $VVV$ channels, distinct from the pattern predicted by in the SM or by new physics manifesting itself via anomalous triple gauge boson couplings. 
In case of an excess over the SM in $VVV$ signatures, observation of the signal in multiple channels would allow one to identify the scenario responsible for it. 

\vspace{2mm}
\hspace{-5mm} {\it (iii)} In connection to the previous point, the signal in different $VVV$ channels depends on the different combinations of the Yukawa corrections $\delta y_u$, $\delta y_d$ and $\delta y_s$. 
In particular, from \eqref{eq:TXS_qqbGzGpGm} it is clear that for the charge $\pm 1$ final states ($W^{\pm}W^{\pm}W^{\mp}$ and $ W^{\pm}ZZ$) the cross-section enhancement is the same for $\delta y_u$ or $\delta y_d$ modifications, and does not allow to distinguish between both. 
In contrast, for 
the neutral final states ($ZW^+W^-$ and $ZZZ$), the different partonic content of up and down quarks in the proton leads to different cross-section enhancements for $\delta y_u$ and $\delta y_d$. 
This is shown explicitly in Table~\ref{Table_1} (which is based on complete calculations, rather than the Golstone boson approximation at high-energies).  
Assuming $\delta y_s =0$, observation of a combined excess in neutral and charge $\pm 1$  $VVV$ channels would allow one to disentangle $\delta y_u$ from $\delta y_d$.   

\vspace{2mm}

\hspace{-5mm} {\it (iv)}  Finally, we note that this observable is different from probes involving on-shell Higgs bosons, since the set of operators that can modify the Higgs boson production and decay patterns is much larger than for triboson.

\begin{table*}[t]
\centering
\begin{tabular}{|c|c|c|c|c|}
\hline
{\bf HL-LHC} & SM & BSM ($Y_d=1$)
& BSM ($Y_u=1$) 
& BSM ($Y_s=1$) \\ \hline
$W^+ W^- W^+$ & 152 fb & 3.6 pb & 3.6 pb & 110 fb \\ \hline
$W^+  W^- W^-$ & 87 fb & 1.5 pb & 1.5 pb  & 110 fb \\ \hline
$Z Z W^+$ & 40 fb & 1.0 pb & 1.0 pb & 31 fb \\ \hline
$Z Z W^-$ & 23 fb & 0.43 pb & 0.43 pb & 31 fb \\ \hline
$Z W^+ W^-$ & 191 fb & 1.5 pb & 2.4 pb & 120 fb\\ \hline
$Z Z Z $ & 16 fb & 0.99 pb & 1.7 pb  & 66 fb\\ \hline
\end{tabular}
\end{table*}
\begin{table}[t]
\centering
\hspace{1mm}
\begin{tabular}{|c|c|c|c|c|}
\hline
{\bf FCC-hh} & SM & BSM ($Y_d=1$)
& BSM ($Y_u=1$) 
& BSM ($Y_s=1$) \\ \hline 
$W^+ W^- W^+$ & 2.35 pb & 290 pb & 290 pb & 16 pb\\ \hline
$W^+  W^- W^-$ & 1.76 pb & 140 pb & 140 pb & 16 pb \\ \hline
$Z Z W^+$ & 756 fb & 74 pb & 74 pb & 4.4 pb \\ \hline
$Z Z W^-$ & 579 fb & 36 pb & 36 pb & 4.4 pb \\ \hline
$Z W^+ W^-$ & 3.93 pb & 94 pb & 150 pb & 12 pb \\ \hline
$Z Z Z $ & 231 fb & 110 pb & 180 pb & 11 pb \\ \hline
\end{tabular}
\label{Table_1}
\caption{Values of different triboson production cross sections for $\sqrt{s} = 14$ TeV LHC (upper table) and $\sqrt{s} = 100$ TeV FCC-hh (lower table) for the SM (computed at NLO in QCD~\cite{Dittmaier:2017bnh,Binoth:2008kt}) and with the addition of the dimension-6 operators from Eq.~\eqref{eq:TH_d6}, with 
$Y_d = 1$ (with $Y_{\neq d} = 0$), $Y_u = 1$ (with $Y_{\neq u} = 0$) and $Y_s = 1$ (with $Y_{\neq s} = 0$), respectively. These latter cross sections are computed at LO. 
}
\end{table}

\section{Triple heavy vector boson channels}
\label{sec:triple}

In this Section we analyse in detail the sensitivity of triple EW gauge boson production to the light quark Yukawa modifications parametrized by $\delta y_u$, $\delta y_d$ and $\delta y_s$. We will focus on the HL-LHC with $\sqrt{s} = 14$~TeV and a future FCC-hh collider with $\sqrt{s} = 100$~TeV. In our analysis we will assume, for simplicity, that only one of these modifications is present at a time. However we note that combining several triboson channels allows one, in principle, to disentangle the different $\delta y_q$.

We begin by showing in Table~\ref{Table_1} the various triboson production cross-sections, $\sigma(p p \to V V V)$, at the LHC and FCC-hh, turning on one BSM Yukawa contribution from Eq.~\eqref{eq:TH_d6} at a time. The cross-sections for the SM are computed at NLO in QCD~\cite{Dittmaier:2017bnh,Binoth:2008kt}. The BSM cross-sections are dominated by the quadratic contribution since the interference between SM and BSM contributions is proportional to the SM Yukawa coupling of the light quarks and thus negligible. We use {\tt MadGraph5$\_$aMC@NLO}~\cite{Alwall:2014hca} using the NNPDF 2.3 PDF set for our simulations. We implemented the relevant BSM interactions from~\eqref{eq:TH_yukawa} in the unitary gauge using {\tt Feynrules} \cite{Alloul:2013bka}.
From Table~\ref{Table_1}, we see that the triboson channel, which presents the largest cross section enhancement 
with respect to the SM, is by far $p p \to Z Z Z$. Given sufficient luminosity it could be the most sensitive channel and, due to the smallness of the signal in the SM, a smoking gun for new physics. Final state leptons reconstructing the $Z$ also provide a clean final state at a hadron collider. However, the smaller cross-section for this channel compared to others particularly due to the small $Z$ boson branching fraction into leptons, will limit considerably the final sensitivity.
On the other hand, the $p p \to W^{\pm} W^{\pm} W^{\mp}$ channel has the largest triboson production cross section, which also makes it key for our sensitivity analysis. 
The $W^+W^-W^\pm$  and $ZZZ$ production will be the two channels we will target in our sensitivity analysis below, bearing in mind that the addition of the remaining channels in Table~\ref{Table_1} would further increase the sensitivity to $\delta y_q$ in triboson processes. 
In the next subsections we consider in turn the $WWW$ and $ZZZ$ channels.

\subsection{$WWW$: same-sign di-lepton final state}
\label{sec:WWWSS}

The same-sign leptonic channel corresponds to the process $pp \to W^{\pm}W^{\pm}W^{\mp} \to  \ell^{\pm} \ell^{\pm} \nu\nu jj$, with $\ell \equiv e,\mu$. We start by considering the recent 13 TeV CMS search for triple gauge boson production with 137~fb$^{-1}$~\cite{CMS:2020gvq,CMS_SLIDES}, that can already be used to put a limit on $\delta y_q$, choosing $\delta y_d$ first as an example. The $p p \to W W W \to \ell^{\pm} \ell^{\pm} \nu \nu j j$~signal cross section as a function of $Y_d$ is approximately given by
\begin{equation}
\label{eq:XS_YD}
\sigma (Y_d) = 7.5\,{\rm fb} + Y_d^2 \times 210 \,{\rm fb} \, ,
\end{equation}
where we have omitted the negligible interference term. The SM cross section is given at NLO in QCD~\cite{Dittmaier:2017bnh}, and we have multiplied the BSM signal cross section obtained from~{\tt MadGraph5$\_$aMC@NLO} by an NLO $k$-factor, $k=1.28$~\cite{Alasfar:2019pmn}. 
According to the CMS analysis, the $\mu^{\pm} \mu^{\pm}$ and $e^{\pm} \mu^{\pm}$ final states in the ``$m_{jj}$-in" category (where the two leading jets in the event reconstruct the $W$ mass, $m_{jj} \sim m_W$) are the most sensitive of the two-lepton same-sign ($2 \ell^{\mathrm{SS}}$) categories (see auxiliary Fig.~24 of Ref.~\cite{CMS:2020gvq}), and we concentrate on those here to obtain a conservative limit. 
The relevant CMS analysis cuts for the $2 \ell^{\mathrm{SS}}$ in this category are
\begin{align}
\label{2L_Cuts}
& p_{T}^{\ell_{1,2}} > 25 \,{\rm GeV} \, ,\,\,\, m_{\ell\ell} > 20 \,{\rm GeV} \, ,\,\,\, m_{j j} \in [65, 95] \,{\rm GeV} \, (``m_{jj}\,{\rm in}") \, ,\,\,\,  \nonumber  \\ 
& E^{{\rm miss}}_T > 45 \,{\rm GeV} \, ,\,  m^{{\rm max}}_T(\ell) > 90 \,{\rm GeV} \, ,
\end{align}
with $\ell_1$ and $\ell_2$ the leading and subleading leptons in $p_T$, and $m^{{\rm max}}_T(\ell)$ defined as the transverse mass built from the missing transverse energy $E^{{\rm miss}}_T$ and the hardest lepton. The cut efficiencies for the BSM ($\epsilon_S$) and the SM ($\epsilon_B$) triboson signals are
\begin{equation}
\label{2L_effs}
\epsilon_S =\,  0.45 \; \quad , \quad 
\epsilon_B = \, 0.27  \, .  
\end{equation}
Note that the CMS analysis is designed to extract the SM triboson signal from the SM background, hence the relatively high $\epsilon_B$ here. The SM background and observed SM signal events for the cut-based CMS ``$m_{jj}$-in" $\mu^{\pm} \mu^{\pm}$ and $e^{\pm} \mu^{\pm}$ selection categories are given in the auxiliary Table 5 of Ref.~\cite{CMS:2020gvq}. For the $e^{\pm} \mu^{\pm}$ category, the expected SM background, expected SM $WWW$ signal and the observed number of events are, respectively, 35.2, 3.3 and 46. For the $\mu^{\pm} \mu^{\pm}$ category these numbers are 24.6, 3.5 and 20. 

By computing the expected ratio of BSM to SM triboson events as a function of $Y_d$ from~\eqref{2L_Cuts} and~\eqref{2L_effs} and normalizing to the CMS expected number of SM triboson events in each signal region, we obtain a $2\sigma$ bound\footnote{Here and in all of the following, the bounds implicitly refer to the magnitude of $\delta y_q$, that is $\delta y_q \equiv |\delta y_q|$.}
 \begin{align}
\begin{split}
 & \delta y_d \lesssim \, 6800 \;\;\; \text{(LHC CMS analysis~\cite{CMS:2020gvq})} \, .
 \end{split}
 \end{align}

\begin{figure}[t]
\centering

\includegraphics[width=0.485\textwidth]{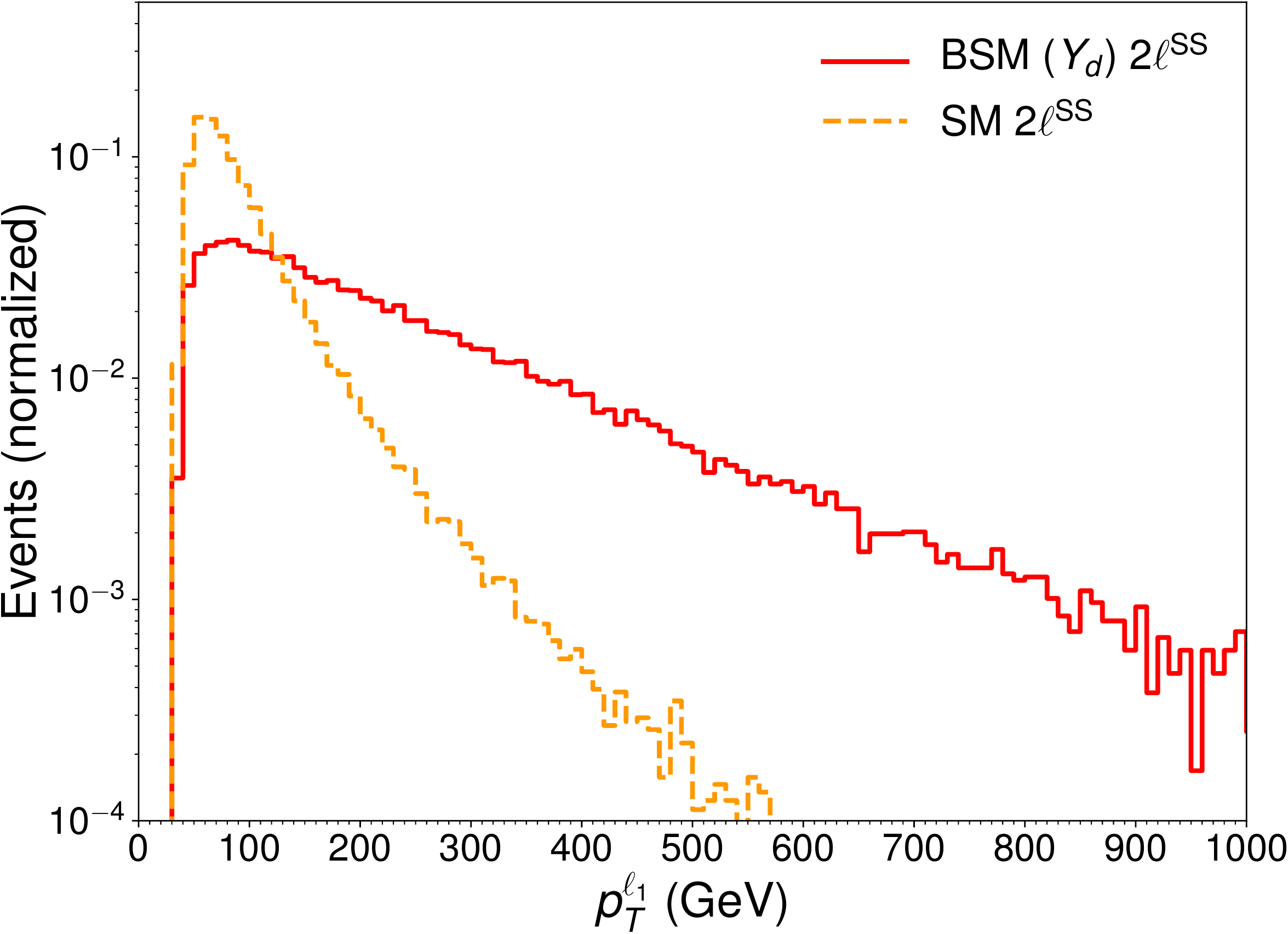} 
\includegraphics[width=0.485\textwidth]{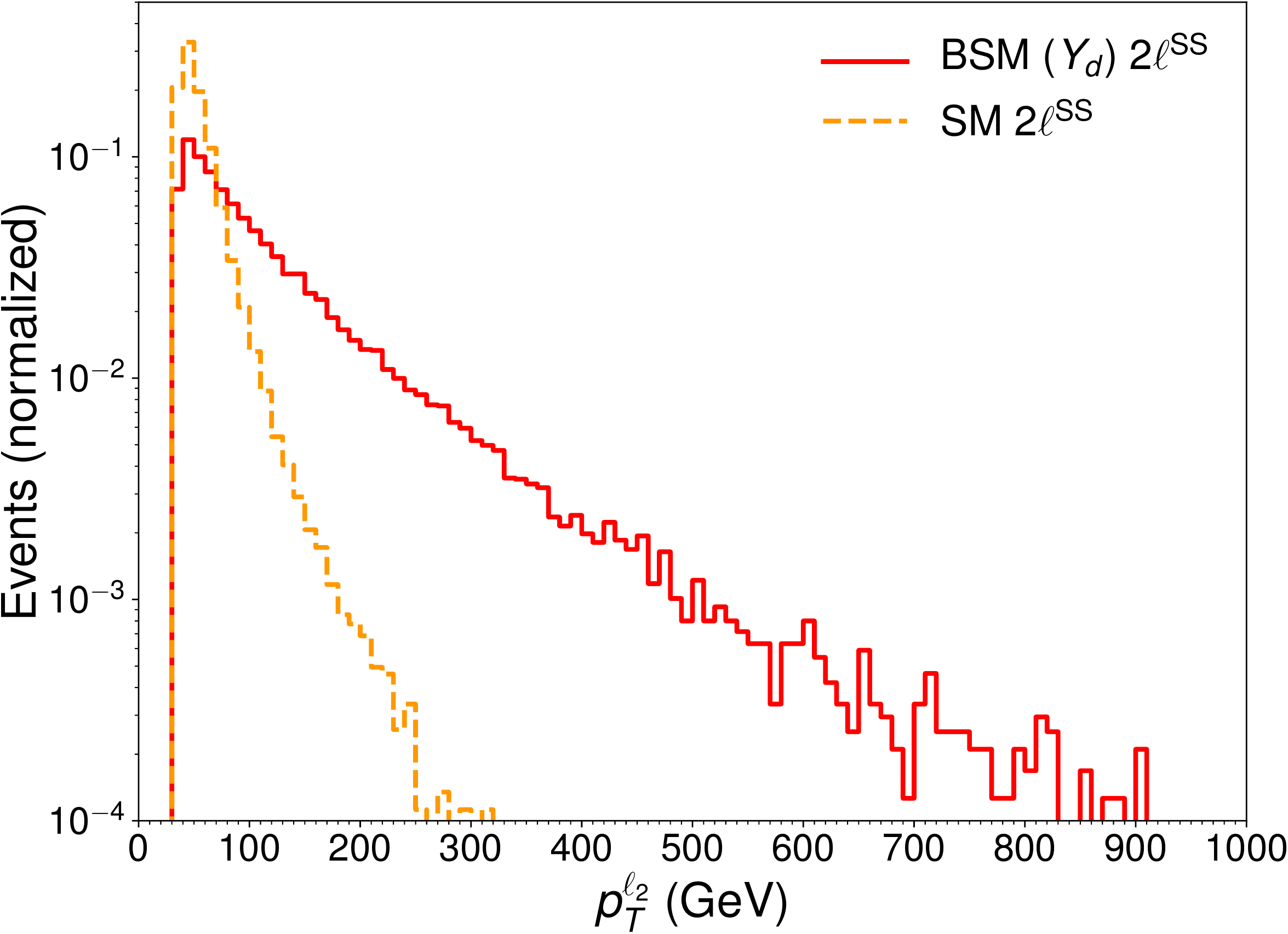} 
\vspace{2mm}

\includegraphics[width=0.485\textwidth]{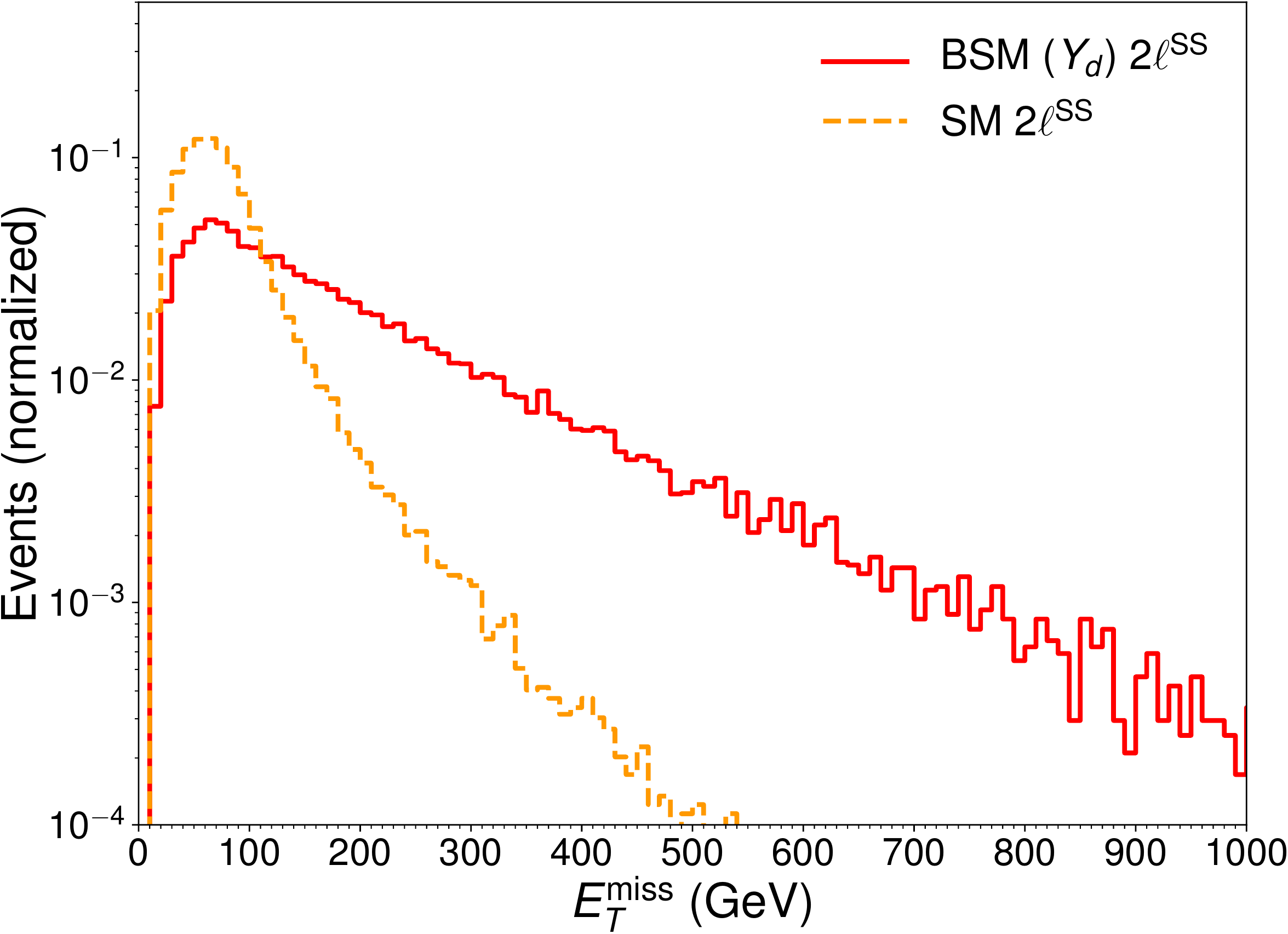} 
\includegraphics[width=0.485\textwidth]{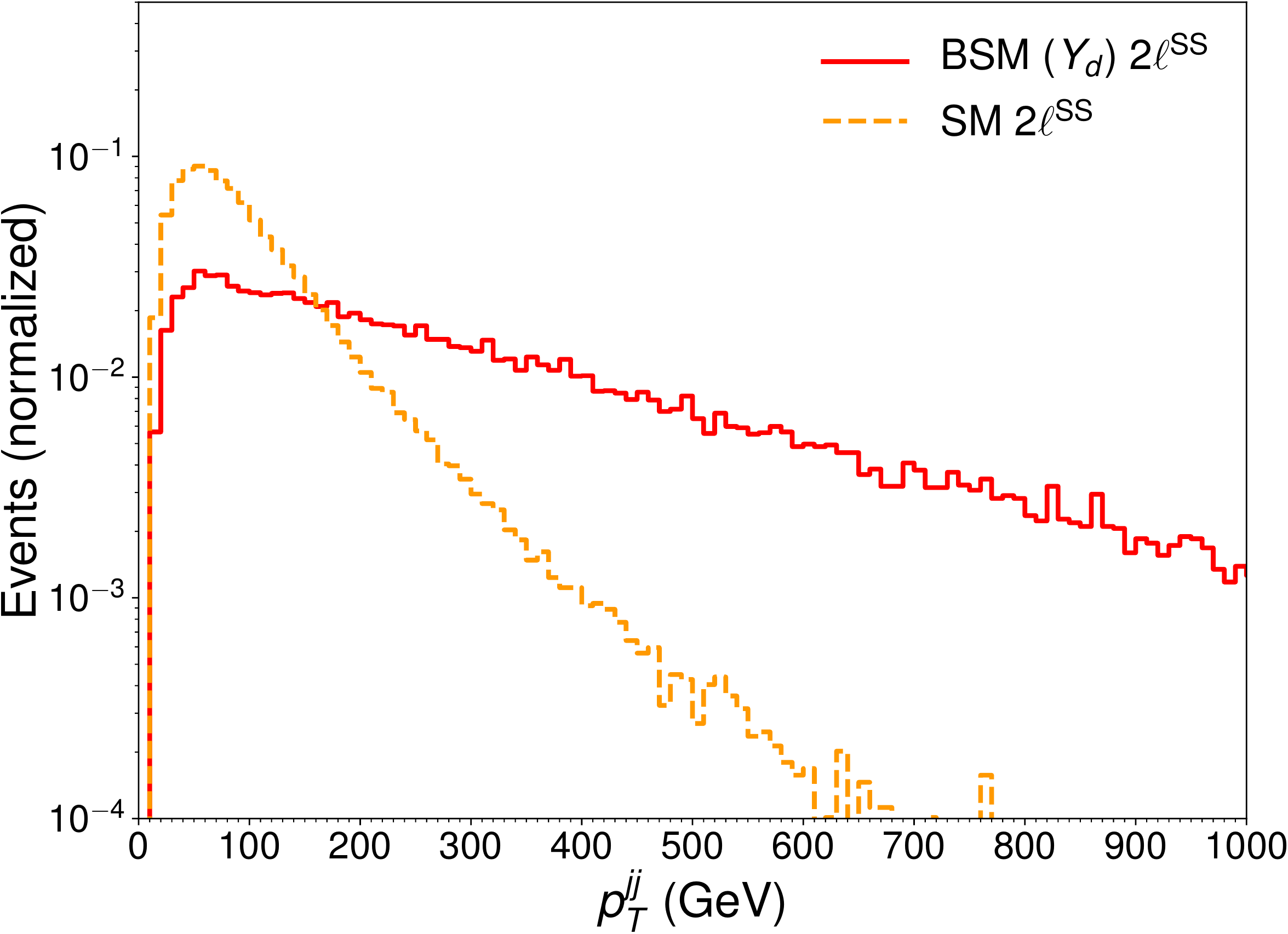} 
\caption{ \small Normalized (to unit area) $WWW$ same-sign di-lepton channel differential distributions for $p_{T}^{\ell_1}$ (top-left), $p_{T}^{\ell_2}$ (top-right), $E^{{\rm miss}}_T$ (bottom-left) and $p_{T}^{jj}$ (bottom-right), for the pure $Y_d$ BSM triboson signal (solid-red lines) and the SM triboson contribution (dashed-yellow lines) at the 14 TeV LHC.}
\label{fig:2lepSS-distributions}
\end{figure} 

It is clear that the current $2 \ell^{\mathrm{SS}}$ CMS analysis is not tailored to search for deviations in the light-quark Yukawa couplings from their SM values. Large improvements in the sensitivity to $\delta y_q$ can be obtained by a more stringent event selection, as illustrated in Fig.~\ref{fig:2lepSS-distributions} for the normalized $p_{T}^{\ell_1}$, $p_{T}^{\ell_2}$, $E^{{\rm miss}}_T$ and $p_{T}^{jj}$ (transverse momentum of the leading di-jet system) distributions for BSM and SM triboson $2 \ell^{\mathrm{SS}}$ signal at the $\sqrt{s} = 14$ TeV LHC.
To illustrate this large potential improvement, we apply the following set of cuts for a $2 \ell^{\mathrm{SS}}$ triboson analysis at the HL-LHC,
\begin{equation}
p_T^{\ell_{1,2}} > 60 \, \text{GeV}\; , \quad E^{\rm miss}_T  > 120 \, \text{GeV} \; ,\quad p_T^{jj}> 120  \, \text{GeV} \; ,\quad |\Delta \eta(\ell_1,\ell_2)|<2 \, ,
\label{LHC_2lSS_Cuts}
\end{equation}
where the two jets ($jj$) come from the reconstructed hadronically-decayed $W$. For an analysis at the FCC-hh, the harder kinematics for the BSM triboson signal as compared to the LHC allows for even tighter cuts,
\begin{equation}
p_T^{\ell_{1,2}} > 100 \, \text{GeV}\; , \quad E^{\rm miss}_T  > 150 \, \text{GeV} \; , \quad p_T^{jj}> 150  \, \text{GeV} \; , \quad |\Delta \eta(\ell_1,\ell_2)|<2 \, .
\label{FCC_2lSS_Cuts}
\end{equation}
 The leptons are considered within the pseudorapidity acceptance $|\eta|<2.5$. A veto on a third lepton with $p_T>10\,$GeV is applied.
 We find the following efficiencies for the $\delta y_d$ BSM ($\epsilon_S$) and SM ($\epsilon_B$) triboson processes,
 \begin{align}
\begin{split}
 & \epsilon_S =\,  0.61 \; \text{(HL-LHC)} \quad  ,\quad \epsilon_S =\,  0.61 \; \text{(FCC-hh)} \, , \\
 & \epsilon_B = \, 0.015 \;  \text{(HL-LHC)} \quad , \quad \epsilon_B =\,  0.0055 \; \text{(FCC-hh)} \, .\\
 \end{split}
 \end{align}
This analysis can be repeated for a BSM contribution with $\delta y_u \neq 0$ (same efficiencies as for $\delta y_d \neq 0$) or $\delta y_s \neq 0$, and for the case of a strange Yukawa modification we obtain an efficiency $\epsilon^{s}_S=0.48$ at the HL-LHC and $\epsilon^{s}_S=0.4$ at the FCC-hh with the cuts~\eqref{LHC_2lSS_Cuts} and~\eqref{FCC_2lSS_Cuts}, respectively. We assume throughout a luminosity of $2 \times 3$ ab$^{-1} = $ 6 ab$^{-1}$ for ATLAS and CMS combined at the HL-LHC and $2 \times 15$ ab$^{-1} = $ 30 ab$^{-1}$ at FCC-hh.  
 
\vspace{1mm}

In the limit of negligible reducible SM background (which we estimate below) to the $2 \ell^{\mathrm{SS}}$ triboson search, the estimated 2$\sigma$ bounds on the Higgs Yukawa couplings to light quarks from the above event selection are
 \begin{align}
\begin{split}
 & \delta y_d \lesssim \, 550 \; \text{(HL-LHC)}\quad , \quad \lesssim \, 63 \;  \text{(FCC-hh)} , \\
 & \delta y_u  \lesssim \, 1100 \;  \text{(HL-LHC)} \quad , \quad \lesssim \, 130 \;  \text{(FCC-hh)},  \\
 & \delta y_s  \lesssim \, 150 \;  \text{(HL-LHC)} \quad , \quad \lesssim \, 15 \; . \text{(FCC-hh)} .
 \end{split}
 \label{SS2Lprojection1}
 \end{align}

We now estimate the effect of the reducible SM backgrounds. We compute the dominant contributions, identified as the production of a $t\bar t$ pair in association with a weak boson ($t \bar{t} W^{\pm}$, $t \bar{t} Z$), as well as the process $p p \to W^{\pm} Z \,jj$ with one of the leptons from the weak boson decays falling out of the detector acceptance. We estimate these processes at NLO in QCD for $t\bar{t}$V ($V=W^\pm,Z$) and LO for $Z\,jj$. A $b$-tagging performance similar to the CMS DeepCSV $b$-tagging algorithm~\cite{Sirunyan:2017ezt} is assumed for both HL-LHC and FCC-hh cases. A veto on the presence of $b$-jets can be applied to reject a large fraction of the $t \bar{t} W^{\pm}$ and $t \bar{t} Z$ events (the corresponding veto efficiency on the BSM signal is neglected). The sum of $t \bar{t} W^{\pm}$ and $t \bar{t} Z$ reducible SM backgrounds then account for an event yield equal to 5\% (21\%) of the irreducible SM background at the HL-LHC (FCC-hh), while the $W^{\pm} Z \,jj$ SM background accounts for 14\% (47\%).
When including the reducible background, the Higgs Yukawa coupling bounds become:
 \begin{align}
\begin{split}
 & \delta y_d \lesssim \, 570 \; \text{(HL-LHC)}\quad , \quad \; \lesssim \, 71 \; \text{(FCC-hh)} , \\
 & \delta y_u  \lesssim \, 1200 \;  \text{(HL-LHC)} \quad , \quad  \lesssim \, 150 \; \text{(FCC-hh)} , \\
& \delta y_s  \lesssim \, 160 \;  \text{(HL-LHC)} \quad , \quad \; \; \lesssim \, 17 \; \text{(FCC-hh)} , 
 \end{split}
 \end{align}
which shows only a very small degradation in sensitivity with respect to~\eqref{SS2Lprojection1}.

 Finally, the sensitivity can be improved by performing shape analyses of relevant kinematic distributions, as those shown in Fig. \ref{fig:2lepSS-distributions}, which can exploit the different kinematic behaviour of BSM signal compared to the SM. In our study we will consider the $p_T$ distribution of the leading lepton, though we note that further improvement of the sensitivity can be obtained by a comprehensive analysis of multiple distributions. We adopt the following binned log-likelihood: 
 \begin{align}
     \Lambda(\delta y_q)= - 2 \sum_i^{\rm bins} \log\frac{L(S_i+B_i,B_i)}{L(B_i,B_i)} \, 
 \end{align}
where $L(\lambda, k)$ is the Poisson distribution with mean $\lambda$ and occurrence $k$ in each bin, and $S_i(B_i)$ is the expected signal (background) yield in  $i^\text{th}$ bin. We use the following binning for the HL-LHC: bins of 10 GeV from 60 GeV to 600 GeV, 50 GeV from 600 GeV to 1 TeV, and an overflow bin for all events above 1 TeV. For the FCC we use bins of 50 GeV from 100 GeV to 1.6 TeV, 100 GeV until 2.4 TeV, and an overflow bin for all events above 2.4 TeV. The resulting expected sensitivities for $\Lambda(\delta y_q)=4$  are found to be
 \begin{align}
\begin{split}
 & \delta y_d \lesssim \, 430 \; \text{(HL-LHC)}\quad , \quad \lesssim \, 36 \; \text{(FCC-hh)}, \\
 & \delta y_u  \lesssim \, 850 \;  \text{(HL-LHC)} \quad , \quad \lesssim \, 71 \; \text{(FCC-hh)} ,\\
 & \delta y_s  \lesssim \, 150 \;  \text{(HL-LHC)} \quad , \quad \lesssim \, 13 \; \text{(FCC-hh)}.
 \end{split}
 \end{align}
This is the improvement one can expect over the cut and count estimate in Eq.~\eqref{SS2Lprojection1} by using the differential information of the lepton $p_T$.    
 
\subsection{$WWW$: three-lepton final state}
\label{sec:WWW3L}

\begin{figure}[h!]
\centering
\includegraphics[width=0.483\textwidth]{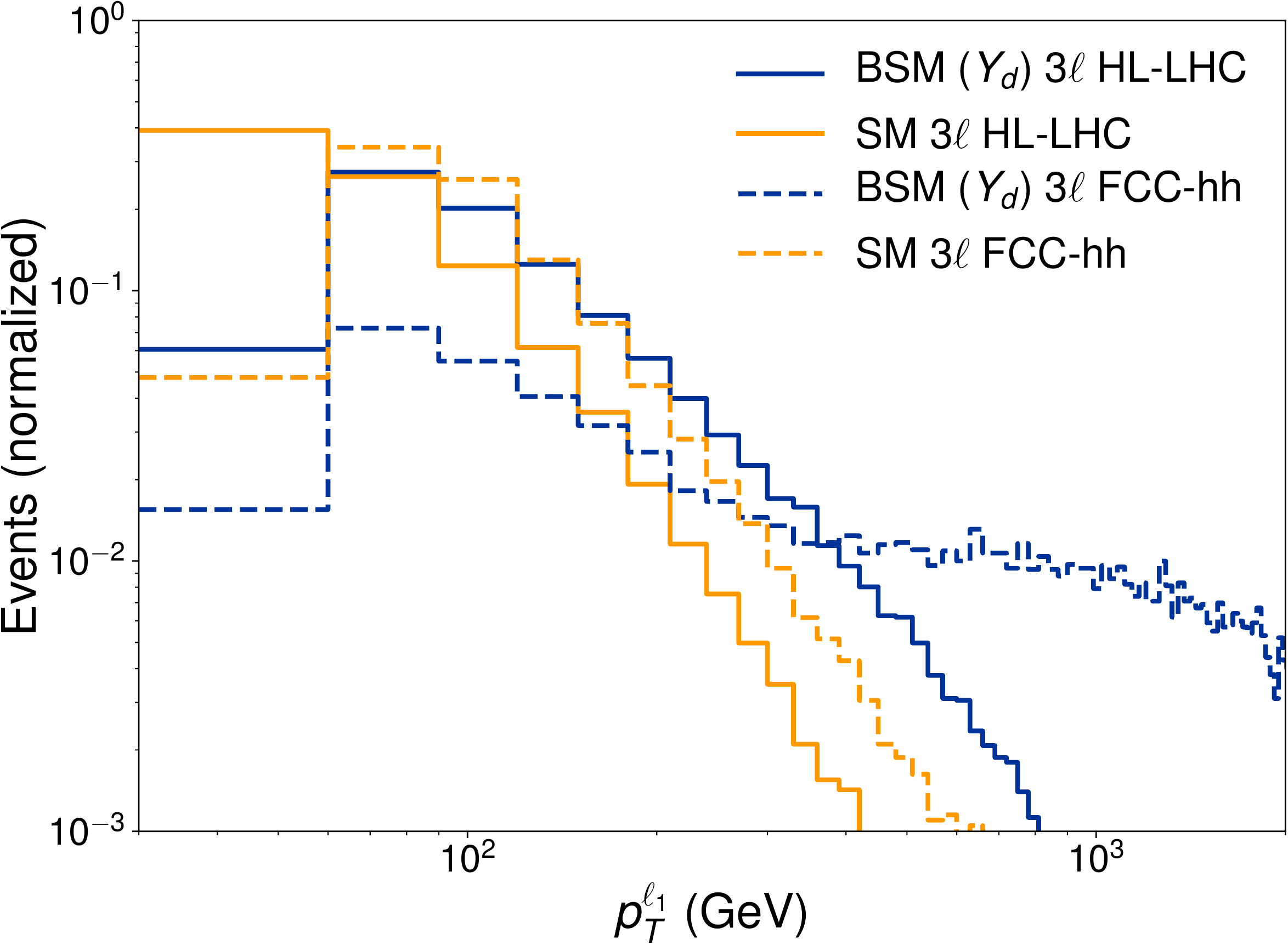} \hspace{2mm}
\includegraphics[width=0.483\textwidth]{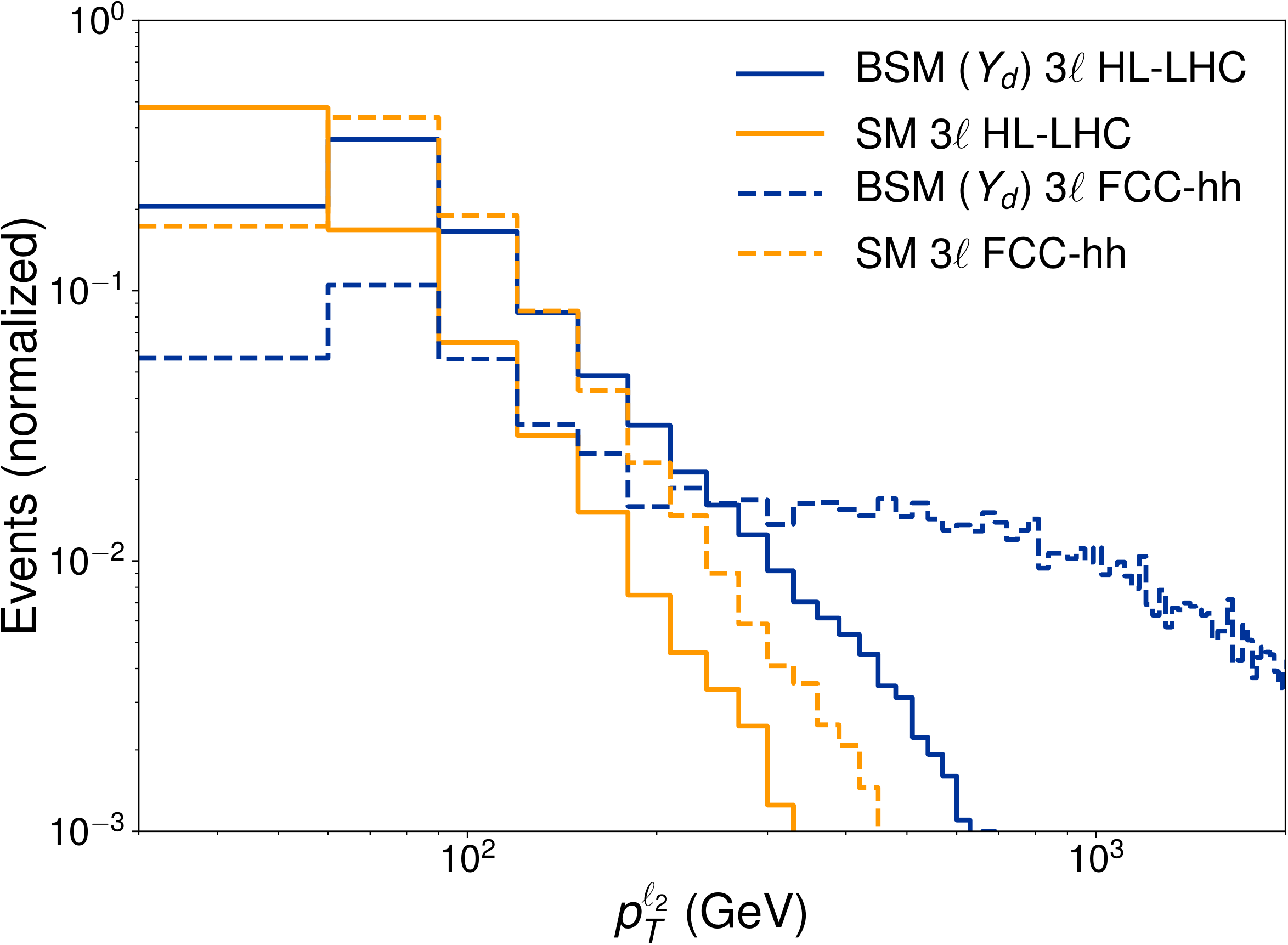} 
\vspace{1mm}

\includegraphics[width=0.483\textwidth]{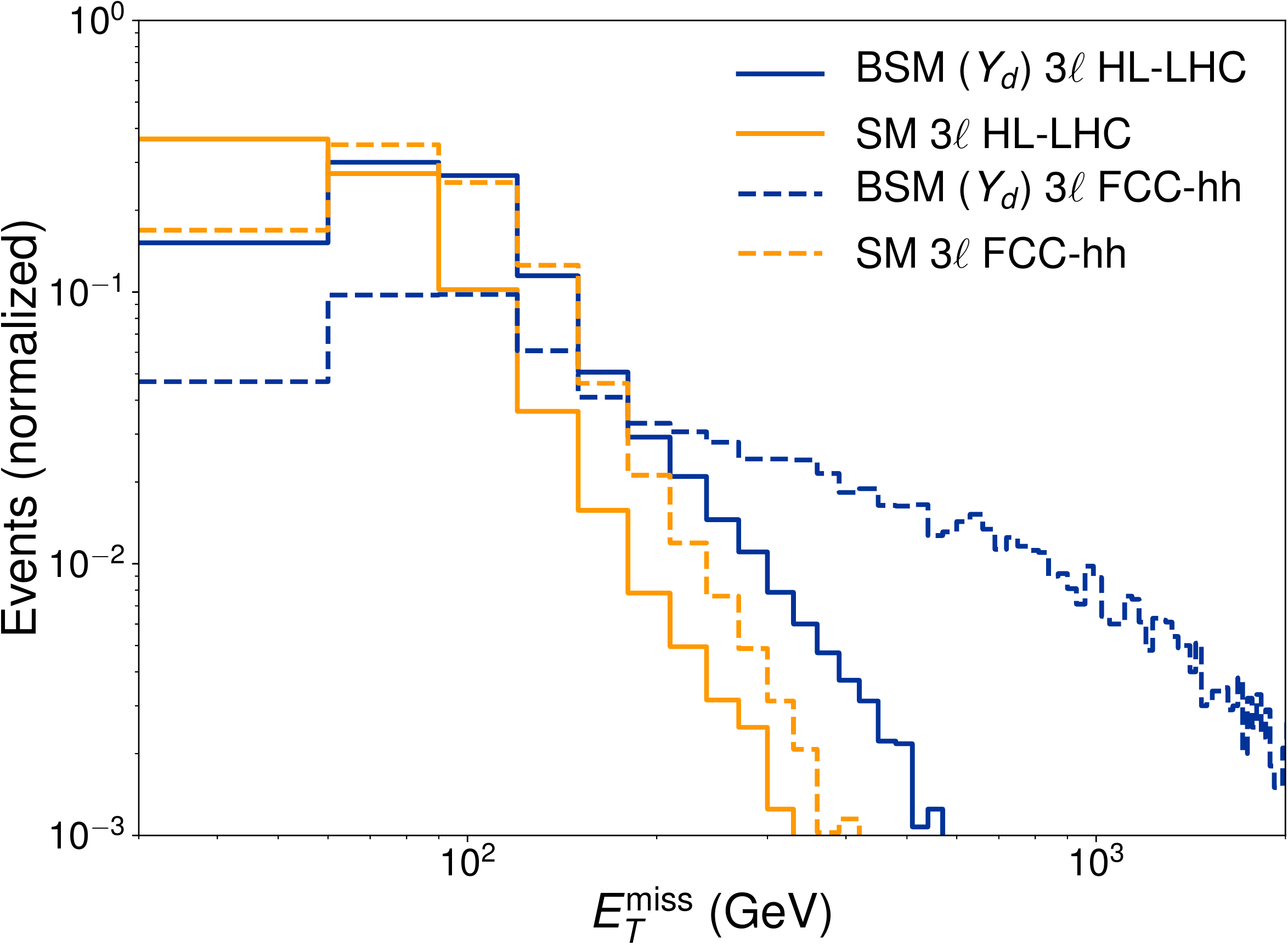} \hspace{2mm}
\includegraphics[width=0.483\textwidth]{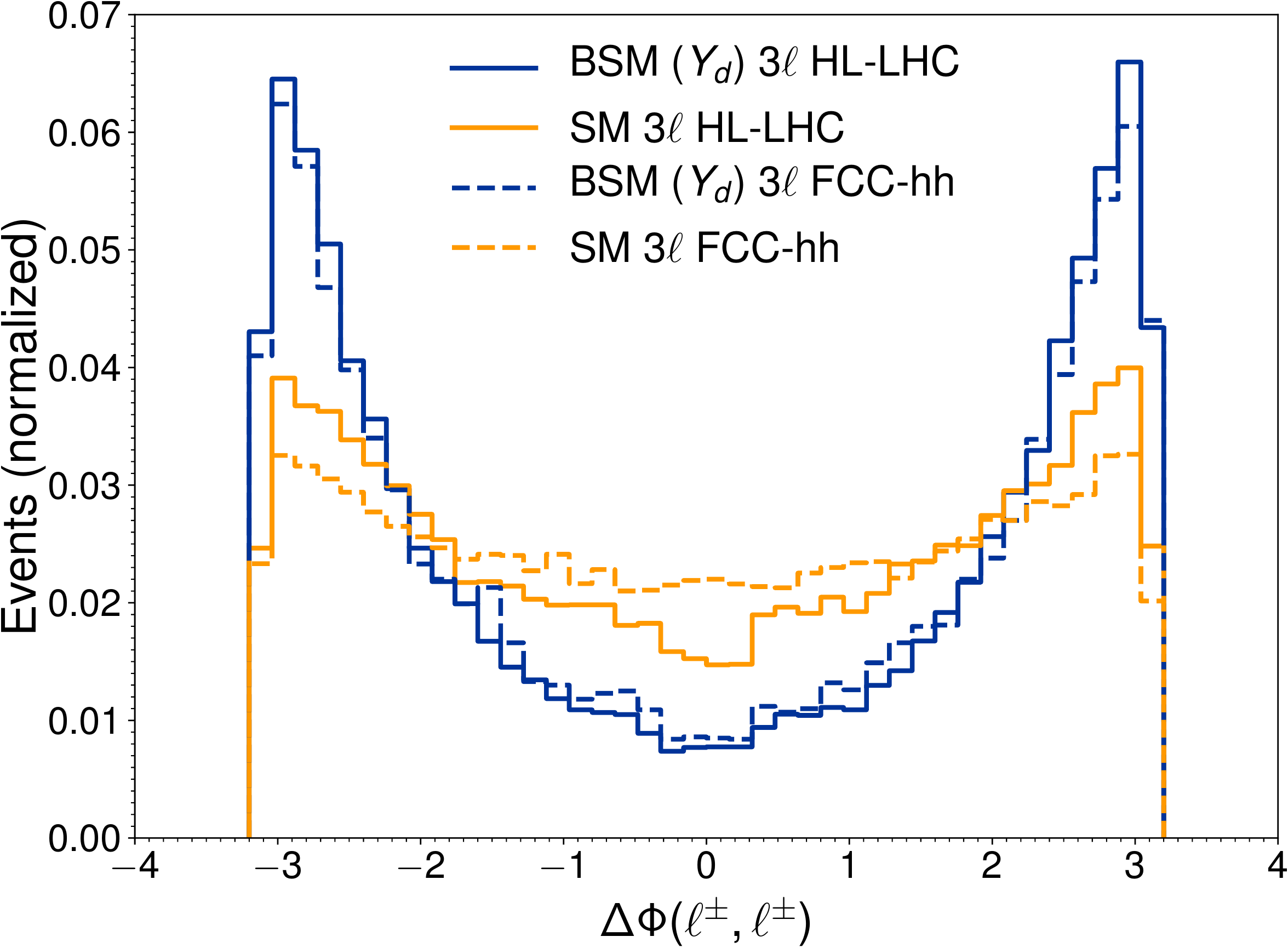}
\caption{ \small 
Normalized (to unit area) $WWW$ tri-lepton channel differential distributions for $p_{T}^{\ell_1}$ (top-left), $p_{T}^{\ell_2}$ (top-right), $E^{{\rm miss}}_T$ (bottom-left) and $\Delta \Phi(\ell^{\pm},\ell^{\pm})$ (bottom-right), for the pure $Y_d$ BSM triboson signal (blue lines) and the SM triboson contribution (yellow lines) at the 14 TeV LHC (solid) and 100 TeV FCC (dashed).
}
\label{fig:3lep-distributions}
\end{figure}

We now analyse the three-lepton channel $pp \to W^{\pm}W^{\pm}W^{\mp} \to  \ell^{\pm} \ell^{\pm} \ell^{\mp} \nu\nu\nu$. The normalized differential distributions for the leading and subleading lepton transverse momenta $p^{\ell_{1,2}}_T$, the missing transverse energy $E^{{\rm miss}}_T$ and the angular separation between the same-sign leptons in the tranverse plane $|\Delta \Phi(\ell^{\pm},\ell^{\pm})|$ of the BSM triboson signal and the SM triboson background are shown in Fig.~\ref{fig:3lep-distributions} for the $\sqrt{s} = 14$ TeV HL-LHC and the FCC-hh. Applying the following selection cuts for the HL-LHC,
\begin{equation}
p_T^{\ell_1} > 70 \, \text{GeV} \, , \, \,p_T^{\ell_2} > 50 \, \text{GeV} \, , \, \,p_T^{\ell_3} > 30 \, \text{GeV} \, , \, \,E^{{\rm miss}}_T  > 80 \, \text{GeV}  \, , \,\, |\Delta \Phi(\ell^{\pm},\ell^{\pm})|>2 \, ,
\label{3L_selection_LHC}
\end{equation}
and a tighter set of cuts for the FCC-hh
\begin{equation}
p_T^{\ell_1} > 150 \, \text{GeV} \, , \,\, p_T^{\ell_2} > 80 \, \text{GeV}\, , \,\,  p_T^{\ell_3} > 50 \, \text{GeV} \, , \, \, E^{{\rm miss}}_T  > 120 \, \text{GeV}  \, , \,\, |\Delta \Phi(\ell^{\pm},\ell^{\pm})|>1.5,
\label{3L_selection_FCC}
\end{equation}
we find the following efficiencies for the $\delta y_{u}$ and $\delta y_{d}$ BSM signal ($\epsilon_S$) and the SM ($\epsilon_B$) triboson processes,
 \begin{align}
\begin{split}
 & \epsilon_S =\,  0.62 \; \text{(HL-LHC)} \quad , \quad  \;\, \epsilon_S =\,  0.50 \; \text{(FCC-hh)} \, ,\\
 & \epsilon_B = \, 0.037 \;  \text{(HL-LHC)} \quad , \quad \epsilon_B = \, 0.014 \; \text{(FCC-hh)} \, .
 \end{split}
 \end{align}
For the BSM signal with $\delta y_s \neq 0$, 
the respective signal efficiencies are $\epsilon^{s}_S=0.6$ at the HL-LHC and $\epsilon^{s}_S=0.16$ at the FCC-hh. 

In the limit of negligible reducible background, the estimated 2$\sigma$ bounds on $\delta y_q$ are

 \begin{align}
\begin{split}
 & \delta y_d \lesssim \, 900 \; \text{(HL-LHC)} \quad , \quad \lesssim \, 120 \; \text{(FCC-hh)} \, ,\\
 & \delta y_u  \lesssim \, 1900 \;  \text{(HL-LHC)} \quad , \quad \lesssim \, 240 \; \text{(FCC-hh)} \, ,\\
 & \delta y_s  \lesssim \, 230 \;  \text{(HL-LHC)} \quad , \quad \;\; \lesssim \, 40 \; \text{(FCC-hh)} \, .
 \label{3L_sensitivity_NoBackground}
 \end{split}
 \end{align}

 The dominant reducible backgrounds consist on the production of a top pair in association with a heavy vector boson, with at least three leptons in the final state.
 We have computed the corresponding $t \bar{t} W^{\pm}$ and  $t \bar{t} Z$ processes at NLO in QCD for HL-LHC and FCC-hh.
 Imposing a $b$-jet veto after the selections~\eqref{3L_selection_LHC} and~\eqref{3L_selection_FCC} 
 reduces the combination of $t \bar{t} W^{\pm}$ and $t \bar{t} Z$ background contributions to a
 5.2\% of the irreducible background for HL-LHC and 3.6\% of the irreducible background for FCC-hh.
 As such, the sensitivity estimates~\eqref{3L_sensitivity_NoBackground} do not change appreciably when the dominant reducible SM backgrounds are included in the analysis.  
 We also note that $ZW^{\pm}W^{\mp}$ and $ZZW^{\pm}$ processes, which could in principle constitute reducible backgrounds for our $W^{\pm} W^{\pm}W^{\mp}$ tri-lepton search, can be made negligible by a fourth lepton veto in combination with a di-lepton $Z$-mass veto. Besides, for $\delta y_q \neq 0$ the $ZW^{\pm}W^{\mp}$ and $ZZW^{\pm}$ processes could also be regarded as a BSM signal so it is conservative to omit them in our tri-lepton analysis.

As discussed in Section~\ref{sec:WWWSS}, performing a binned shape analysis may significantly improve the sensitivity with respect to the cut and count selection described above. For example, using the leading lepton $p_T$ differential distribution, as shown in Fig.~\ref{fig:3lep-distributions}, gives the following $2\sigma$ projected bounds from the three-lepton channel alone:
 \begin{align}
\begin{split}
 & \delta y_d \lesssim \, 840 \; \text{(HL-LHC)} \quad , \quad \lesssim \, 54 \; \text{(FCC-hh)} \, ,\\
 & \delta y_u  \lesssim \, 1700 \;  \text{(HL-LHC)}  \quad , \quad \lesssim \, 110 \; \text{(FCC-hh)} \, ,\\
 & \delta y_s  \lesssim \, 230 \;  \text{(HL-LHC)} \quad , \quad \;\; \lesssim \, 33 \; \text{(FCC-hh)} \, .
 \end{split}
 \end{align}
Here, we use the following binning for the HL-LHC: bins of 10 GeV from 70 GeV to 600 GeV, 50 GeV from 600 GeV to 900 GeV, and an overflow bin for all events above 900 GeV. For the FCC we use bins of 10 GeV from 150 GeV to 200 GeV, 50 GeV from 200 GeV to 1.5 TeV, 100 GeV from 1.5 TeV to 2.2 TeV, and an overflow bin for all events above 2.2 TeV.  
 
Finally, we note that the total $WWW$ sensitivity would benefit from combining its various decay channels. For example a combination of the shape analyses for the same-sign di-lepton and three-lepton channels, neglecting correlations, yields the improved bounds 
\begin{align}
\begin{split}
 & \delta y_d \lesssim \, 420 \; \text{(HL-LHC)}\quad , \quad \lesssim \, 34 \; \text{(FCC-hh)} \, ,  \\
 & \delta y_u  \lesssim \, 830 \;  \text{(HL-LHC)} \quad , \quad  \lesssim \, 68 \; \text{(FCC-hh)} \, , \\
 & \delta y_s  \lesssim \, 140 \;  \text{(HL-LHC)} \quad , \quad  \; \lesssim \, 13 \; \text{(FCC-hh)} \, .
 \end{split}
 \end{align}
These sensitivities are clearly dominated by the signal in the same-sign dilepton final state.
 
\subsection{$ZZZ$: four-lepton final state} 

We move to discussing the sensitivity of the $ZZZ$ channel to light Yukawas.
As discussed previously, combining that with the sensitivity analyses of the $WWW$ channel from Sections~\ref{sec:WWWSS} and~\ref{sec:WWW3L} in principle allows one to disentangle the effects of $\delta y_u$ and $\delta y_d$. In addition, the cross section values of Table~\ref{Table_1} indicate that $ZZZ$ production could at the same time yield strong sensitivity to the presence of $\delta y_s \neq 0$. 

Regarding possible $ZZZ$ decay channels, we note that the $6 \ell$ final state, despite being the cleanest channel, suffers from too low a cross-section and thus it does not allow to obtain competitive limits. We then focus here on the $4\ell$ final states: $p p \to ZZZ \to 4\ell + 2\nu$ and $p p \to ZZZ \to 4\ell + 2j$. In the following we shall perform a naive estimate of the sensitivity to $\delta y_q$ in both channels. 

\subsubsection{$4\ell + E^{\mathrm{miss}}_T$}

The $4\ell + 2\nu$ decay channel of the $ZZZ$ triboson process has the advantage of being easy to disentangle from the dominant reducible SM background, $ p p \to Z Z \to 4 \ell$, due to the presence of $E^{\mathrm{miss}}_T$ from the neutrinos in the case of the signal. The two relevant irreducible SM backgrounds for our BSM process are the triboson processes $ZZZ$ and $WWZ$. The latter becomes very suppressed by requiring two same-flavour lepton pairs reconstructing $Z$-masses, i.e. $\left|m_{Z} - m_{\ell\ell} \right| < 10$ GeV for each lepton pair. Similarly, the reducible SM backgrounds $t {\bar t} Z$, $t W Z$ are suppressed to a negligible level by this requirement in combination with a $b$-jet veto~\cite{CMS:2020gvq}. For the $ZZ$ reducible background, the inclusive cross section after the $Z\to \ell\ell$ decays is $\sim 75$ fb~\cite{Aaboud:2017rwm}, much larger than that of the irreducible SM backgrounds. The presence of $E^{\mathrm{miss}}_T$ for $p p \to ZZ \to 4\ell$ is however due to detector resolution and potential mismeasurements, and thus it is expected to be very small above a certain $E^{\mathrm{miss}}_T$ range.

\begin{figure}[h]
\centering
\includegraphics[width=0.48\textwidth]{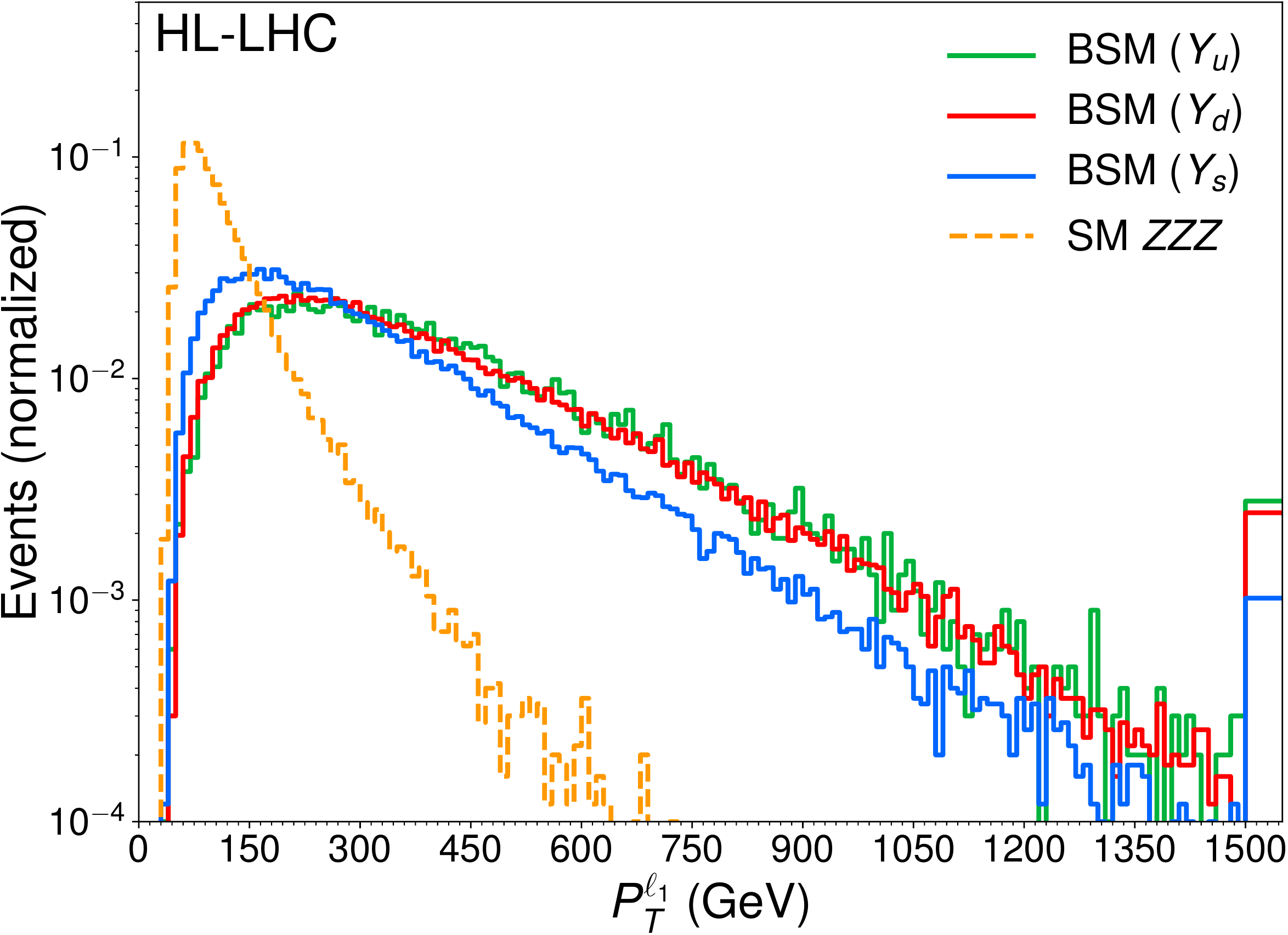} 
\includegraphics[width=0.48\textwidth]{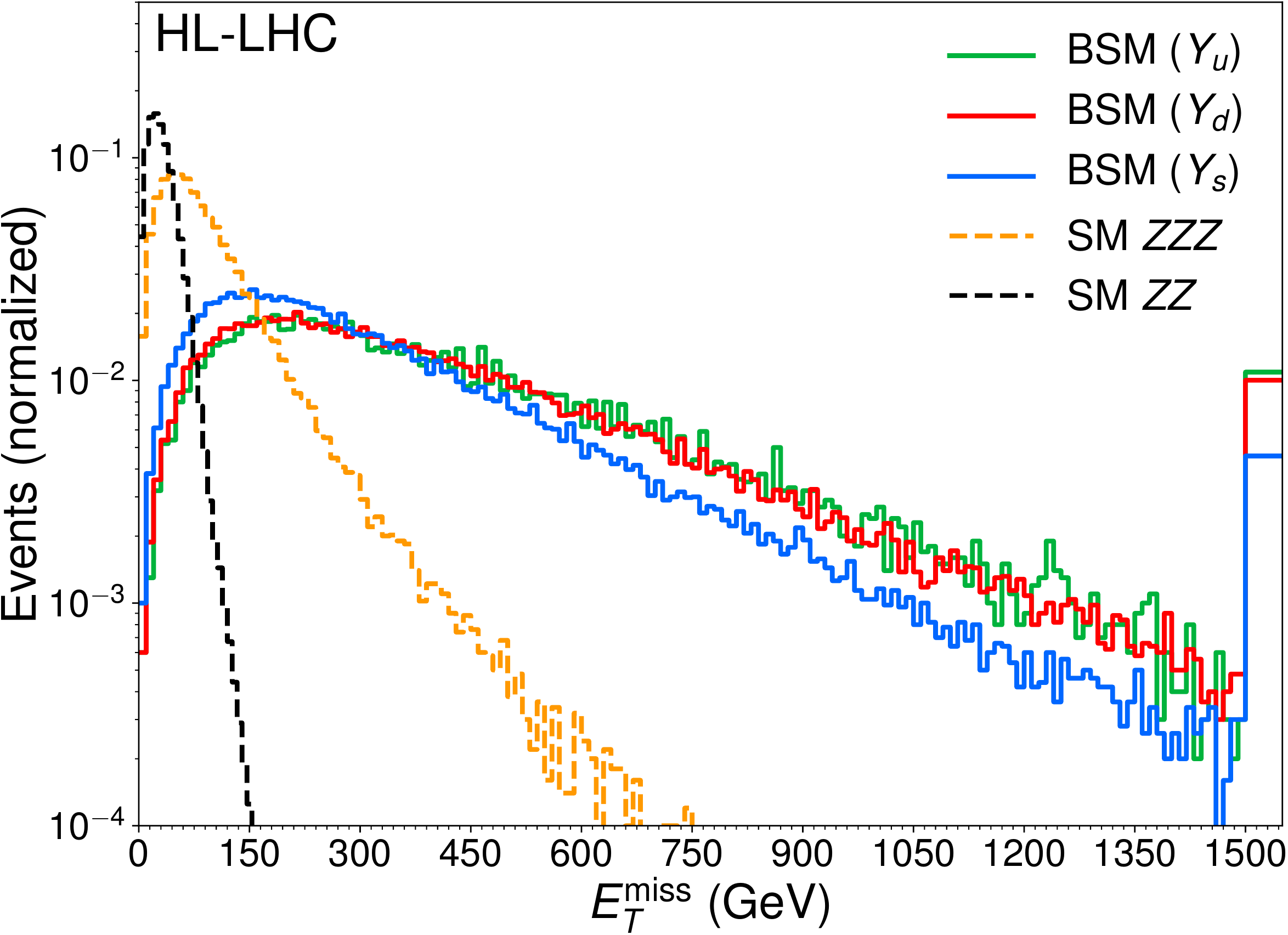} 
\vspace{2mm}

\includegraphics[width=0.48\textwidth]{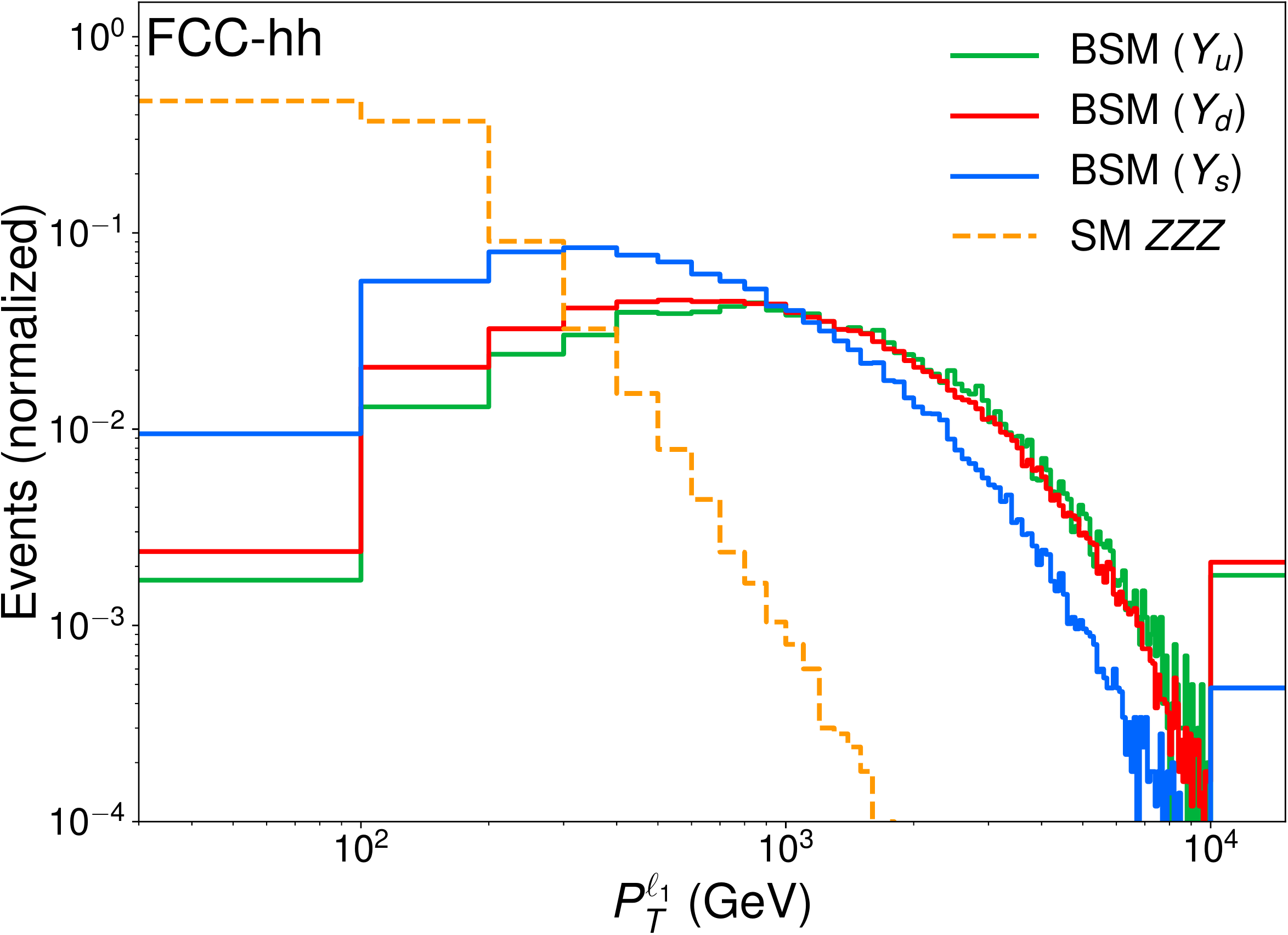} 
\includegraphics[width=0.48\textwidth]{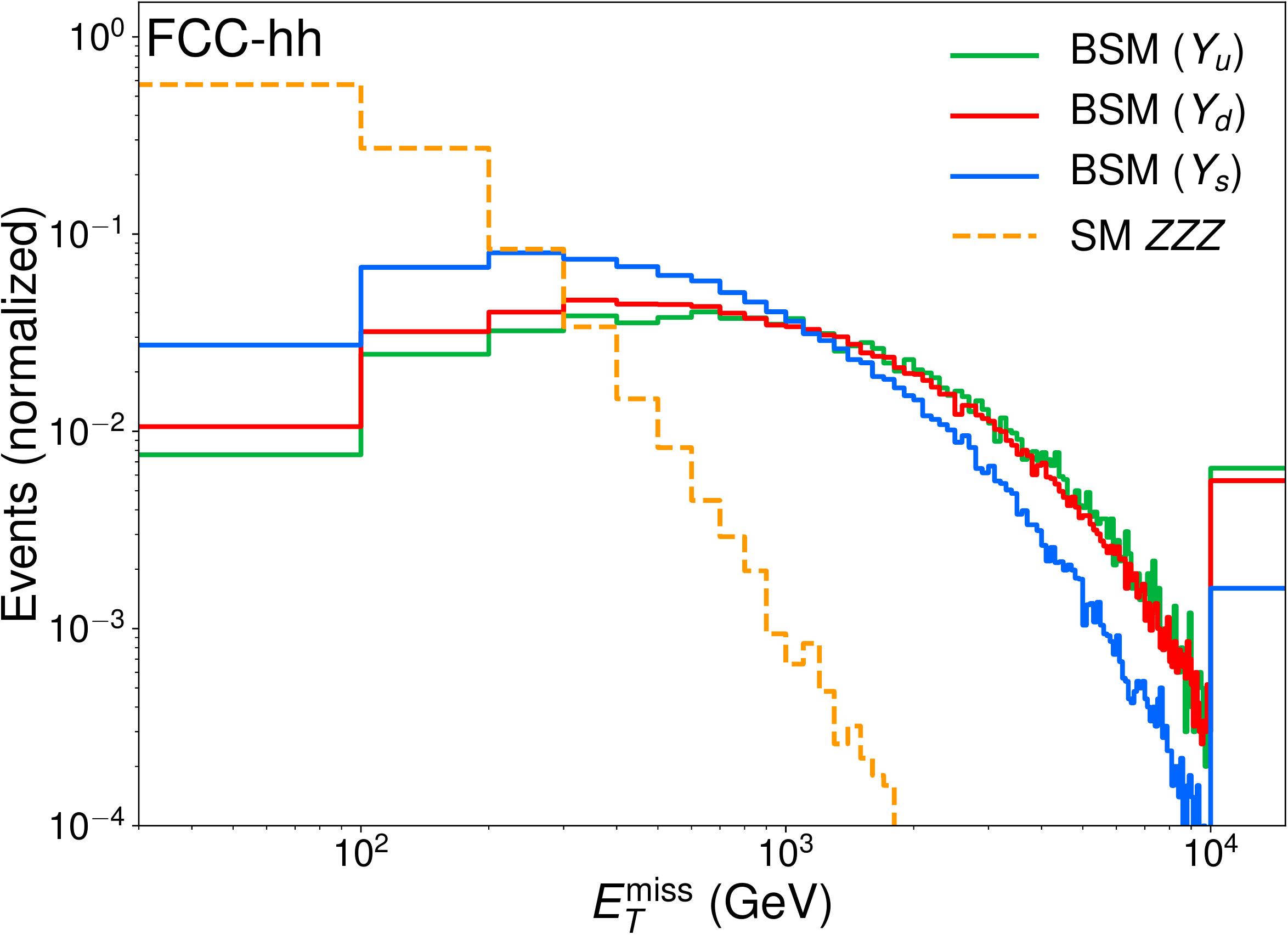} 

\caption{\small 
Normalized (to unit area) $p p \to Z Z Z \to 4 \ell + 2 \nu$ differential distributions for $p_{T}^{\ell_1}$ (left) and $E^{{\rm miss}}_T$ (right) for the pure $Y_d$ BSM triboson signal (solid-red), pure $Y_u$ BSM triboson signal (solid-), pure $Y_s$ BSM triboson signal (solid-blue), and the SM triboson contribution (dashed-yellow) at the 14 TeV LHC (top) and 100 TeV FCC (bottom). For HL-LHC $E^{{\rm miss}}_T$ (top-right) we also include the $Z Z$ reducible SM background as a dashed-black line (see text for details). In each plot, the last bin corresponds to the overflow bin.}
\label{fig:ZZZ}
\end{figure}

In the following, we consider events for which the following initial selection cuts are applied
\begin{equation}
p_T^{\ell_{1,2}} > 25 \text{ GeV} \, , \, p_T^{\ell_{3,4}} > 10 \text{ GeV} \, , \, \left| \eta_{\ell} \right| < 2.5 \, , \, \Delta R_{\ell\ell} > 0.1 \, , \, \left|m_{Z} - m_{\ell\ell} \right| < 10 \text{ GeV} \, . 
\label{4L_MET_Selection_Cuts}
\end{equation}
The $p p \to Z Z Z \to 4 \ell + 2 \nu$~cross section computed at LO for $\sqrt{s} = 14$ TeV LHC with these selection cuts is given by
\begin{eqnarray}
\label{eq:XS_YD2}
\sigma (Y_u) &=& 0.013\,{\rm fb} + Y_u^2 \times 3.0 \,{\rm fb} \,, \nonumber \\
\sigma (Y_d) &=& 0.013\,{\rm fb} + Y_d^2 \times 1.8 \,{\rm fb} \,, \\
\sigma (Y_s) &=& 0.013\,{\rm fb} + Y_s^2 \times 0.14 \,{\rm fb} \, \nonumber ,
\end{eqnarray}
respectively for $\delta y_u \neq 0$, $\delta y_d \neq 0$ and $\delta y_s \neq 0$. In Fig.~\ref{fig:ZZZ} (top) we show the normalized $p_T^{\ell_{1}}$ and $E^{\mathrm{miss}}_T$ distributions for the BSM signal and SM triboson $ZZZ$ background, for the LHC with $\sqrt{s} = 14$ TeV.  We also include in the plot for comparison the expected $E^{\mathrm{miss}}_T$ distribution for the $Z Z$ reducible SM background after the selection~\eqref{4L_MET_Selection_Cuts} for $\sqrt{s} = 13$ TeV LHC, as given in~\cite{CMS_SLIDES} (and then normalized). This distribution can be accurately fitted by an exponentially decreasing function for $E^{\mathrm{miss}}_T \gtrsim 50$ GeV. Thus, an $E^{\mathrm{miss}}_T$ cut greatly suppresses the $ZZ$ reducible background, enhancing at the same time the sensitivity to $\delta y_d$ with respect to~the one obtained directly from~\eqref{eq:XS_YD2}. 
For the HL-LHC sensitivity estimate we select events with $E^{\mathrm{miss}}_T > 200$ GeV. This selection yields a BSM signal efficiency $\epsilon_S = 0.74$ for $\delta y_d \neq 0$, and a SM $ZZZ$ triboson background efficiency $\epsilon_B = 0.09$. For $\delta y_s \neq 0$, the BSM signal efficiency is $\epsilon_S = 0.64$. We also estimate the corresponding efficiency $\epsilon_{ZZ}$ for the reducible SM background via the exponential fit to the $Z Z$ $E^{\mathrm{miss}}_T$ distribution, finding $\epsilon_{ZZ} \sim 1.3 \times 10^{-5}$, which renders it subdominant with respect to the irreducible $ZZZ$ background. 

For FCC-hh, the $p p \to Z Z Z \to 4 \ell + 2 \nu$ cross section at LO reads
\begin{eqnarray}
\label{eq:XS_YD2_FCC}
\sigma (Y_u) &=& 0.11\,{\rm fb} + Y_u^2 \times 340 \,{\rm fb} \,, \nonumber \\
\sigma (Y_d) &=& 0.11\,{\rm fb} + Y_d^2 \times 220 \,{\rm fb} \,, \\
\sigma (Y_s) &=& 0.11\,{\rm fb} + Y_s^2 \times 26 \,{\rm fb}\, \nonumber ,
\end{eqnarray}
with the same basic cuts as for the HL-LHC analysis except for the $\Delta R_{\ell\ell}$ cut, which we set to $\Delta R_{\ell\ell} > 0.01$. The normalized distributions for $p_T^{\ell_{1}}$ and $E^{\mathrm{miss}}_T$ in this case are shown in Fig.~\ref{fig:ZZZ} (bottom). For FCC-hh we set the signal selection cut $E_T^\text{miss} > 500$ GeV.

In order to derive sensitivity projections, we use an NLO $k$-factor of $k=1.28$~\cite{Alasfar:2019pmn}
for the BSM signal, both for LHC and FCC-hh. For the $ZZZ$ background, to normalise to the NLO cross-sections of Table~\ref{Table_1}, we use $k=1.55$ ($1.67$) for HL-LHC (FCC-hh). The resulting projected $2\sigma$ sensitivities after signal selection are found to be 
\begin{align}
\label{Bounds_ZZZ_4lMET_Cut}
&\delta y_d \lesssim 1700 \quad \text{(HL-LHC)} \quad , \quad \lesssim 120 \quad \text{(FCC-hh)} \, , \nonumber \\
&\delta y_u \lesssim 2600 \quad \text{(HL-LHC)} \quad , \quad \lesssim 190 \quad \text{(FCC-hh)} \, , \\
&\delta y_s \lesssim 340 \quad \text{(HL-LHC)} \quad , \quad \lesssim 19  \quad \text{(FCC-hh)} \, \nonumber .
\end{align}
Applying instead a shape analysis to the $E_T^{\rm miss}$ differential distribution via a binned log-likelihood, as described in Section~\ref{sec:WWWSS}, gives the projected sensitivities
\begin{align}
\label{Bounds_ZZZ_4lMET_Shape}
&\delta y_d \lesssim 1500 \quad \text{(HL-LHC)} \quad , \quad \lesssim 65 \quad \text{(FCC-hh)} \, ,\nonumber \\
&\delta y_u \lesssim 2300 \quad \text{(HL-LHC)} \quad , \quad \lesssim 100 \quad \text{(FCC-hh)} \, , \\
&\delta y_s \lesssim 300 \quad \text{(HL-LHC)} \quad , \quad \lesssim 12 \quad \text{(FCC-hh)} \, .\nonumber
\end{align}
We used the following binning of $E_T^{\rm miss}$ for the HL-LHC: bins of 10 GeV from 0 GeV to 800 GeV, 50 GeV from 800 GeV to 1 TeV, and an overflow bin for all events above 1 TeV. For the FCC-hh we use bins of 100 GeV from 0 GeV to 2 TeV, and an overflow bin for all events above 2 TeV.

\subsubsection{$4\ell + 2j$}

Finally we consider the $ZZZ$ sub-channel in which one $Z$-boson decays hadronically, yielding a $4\ell + 2j$ final state. This benefits from more statistics, yet has less clean backgrounds than purely leptonic final states as studied in the previous section: the $4\ell + 2j$ final state has a factor $\sim 3$ higher cross section than the $4\ell +  E^{\mathrm{miss}}_T$ one, but is more difficult to disentangle from the dominant SM reducible background, $ZZ$. 
In fact, as opposed to the previous scenario, the dominant background in this case is given by the SM $ZZ$+jets production. This contribution is estimated at LO in QCD with two partons in the final state ($ZZ\,jj$). In order to distinguish our BSM signal from this background, we first apply a minimal set of cuts, aimed at discarding the events where the two leading jets do not reconstruct a hadronically decaying $Z$-boson. We then exploit the kinematical properties of the BSM signal, which is characterized by harder final particles, by applying a cut on the leading lepton $p_{T}^{\ell_1}>$ 150 GeV (300 GeV) for HL-LHC (FCC-hh). 
The set of cuts we apply is
\begin{equation}
p_T^{\ell_1} > 150 \, (300) \text{ GeV} \, , \, p_T^{\ell_{2,3,4}} > 25 \text{ GeV} \, , \, p_T^{j} > 30 \text{ GeV}  \, , \, M_{jj} \in [81,101] \text{ GeV} ,
\end{equation}
where $M_{jj}$ denotes the invariant mass of the two leading jets. The resulting 2$\sigma$ projected bounds for our cut and count analysis are
\begin{align}
\begin{split}
 & \delta y_d \lesssim \, 1800 \; \text{(HL-LHC)} \quad , \quad \lesssim \, 170 \; \text{(FCC-hh)} \, ,\\
 & \delta y_u  \lesssim \, 2700 \;  \text{(HL-LHC)}  \quad , \quad \lesssim \, 260 \; \text{(FCC-hh)} \, ,\\
 & \delta y_s  \lesssim \,380 \;  \text{(HL-LHC)} \quad , \quad \;\; \lesssim \, 27 \; \text{(FCC-hh)} \, .
 \end{split}
 \end{align}
Performing a shape analysis on the leading lepton $p_T$ distribution, as described in Section~\ref{sec:WWWSS}, gives the improved limits
 \begin{align}
\begin{split}
 & \delta y_d \lesssim \, 1300 \; \text{(HL-LHC)} \quad , \quad \lesssim \, 93 \; \text{(FCC-hh)} \, ,\\
 & \delta y_u  \lesssim \, 1800 \;  \text{(HL-LHC)}  \quad , \quad \lesssim \, 140 \; \text{(FCC-hh)} \, ,\\
 & \delta y_s  \lesssim \,290 \;  \text{(HL-LHC)} \quad , \quad \;\; \lesssim \, 16 \; \text{(FCC-hh)} \, .
 \end{split}\label{eq:shape-4l2j}
 \end{align}
Here, we use the following binning of $p_T^{\ell}$ for both the HL-LHC and FCC: bins of 10 GeV from 40 GeV to 300 GeV, 50 GeV from 300 GeV to 600 GeV, and an overflow bin for all events above 600 GeV. 

Since the two $ZZZ$ channels, $4\ell+E_T^{\rm miss}$ and $4\ell+2j$, provide similar sensitivities in Eqs.~\eqref{Bounds_ZZZ_4lMET_Shape} and \eqref{eq:shape-4l2j}, 
a combination of the shape analyses for the two channels, neglecting correlations, improves the sensitivity of the $ZZZ$ triboson process,
\begin{align}
\begin{split}
 & \delta y_d \lesssim \, 1100 \; \text{(HL-LHC)} \quad , \quad \lesssim \, 60 \; \text{(FCC-hh)} \, ,\\
 & \delta y_u  \lesssim \, 1600 \;  \text{(HL-LHC)}  \quad , \quad \lesssim \, 92 \; \text{(FCC-hh)} \, ,\\
 & \delta y_s  \lesssim \,250 \;  \text{(HL-LHC)} \quad , \quad \;\; \lesssim \, 11 \; \text{(FCC-hh)} \, .
 \end{split}\label{eq:shape-ZZZcombine}
 \end{align}
By comparing these results with those in Sections~\ref{sec:WWWSS} and~\ref{sec:WWW3L}, we see that the $WWW$ $2\ell$ same-sign and $3\ell$ analyses are more sensitive in general due to their higher statistics. Indeed, for $\delta y_{d} \neq 0$ the cross section for $p p \to W^{\pm}W^{\pm}W^{\mp} \to \ell^{\pm}\ell^{\pm} + 2\nu + 2j$ is approximately $100$ times larger than that of $p p \to Z Z Z \to 4 \ell + 2 \nu$. Yet, due to the larger cross section enhancement from LHC to FCC-hh in the case of $ZZZ$ with respect to $WWW$ (see Table~\ref{Table_1}), the sensitivity to $\delta y_{q} \neq 0$ in $ZZZ$ channels at FCC-hh becomes competitive with that of $WWW$, particularly for the strange Yukawa coupling. 

\section{Comparison with other constraints} 
\label{sec:comparison}

The sensitivity of the triboson analysis established in the previous Sections should be compared to the sensitivity of other existing probes of the light quark Yukawa couplings.  

A change in the Higgs decay width into light quarks has an indirect effect on the event rate in other decay channels measured by the LHC collaborations. 
In particular, assuming only one Yukawa coupling is modified at a time,
the total Higgs signal strength normalized to the SM value is given by the expression  
\beq
\label{eq:mu}
\mu =  \frac{1}{1  + \sum_{q}( 2 \delta y_q +  \delta y_q^2) {\rm Br} (h \to qq)_{\rm SM}} \,. 
\eeq 
The above holds when the total Higgs production cross section is not significantly affected by the enhanced $q \bar q \to h$ $q g \to q h$ modes,
which is a good approximation for $\delta y_{u} \lesssim 1000$, $\delta y_{d} \lesssim 500$.
Given the ${\rm Br} (h \to qq)_{\rm SM}$ in Table~\ref{tab:ysm}, 
for $q=u,d,s$ the effect is observable only for $|\delta y_q| \gg 1$, and then the linear term in $ \delta y_q$ in Eq.~\eqref{eq:mu} is subleading compared to the quadratic one.
Note that in this regime the Higgs signal strength is always {\it suppressed}, $\mu < 1$. 
The HL-LHC is expected to measure the total Higgs signal strength with an error of order 2-3\%~\cite{Cepeda:2019klc}.
A future measurement $\mu = 1.00 \pm 0.03$ would set the following bounds on the Yukawa couplings:\footnote{As before, the left-hand-sides of all the $\delta y_q$ bounds should be read as $|\delta y_q|$, but we omit the absolute value sign to simplify the notation.}
\beq
\label{eq:mu-future}
\quad  \delta y_d \lesssim 340,
\quad  \delta y_u \lesssim 700, 
\quad  \delta y_s \lesssim 17   \;\;\;\;\; \text{(HL-LHC)} \, .  
\eeq 
If the effect of the enhanced $q \bar q \to h$ is taken into account, 
the first two limits are slightly relaxed: $\delta y_d \lesssim 360$, $\delta y_u \lesssim 780$. 
Using the most recent measurements from CMS ($\mu = 1.02^{+0.07}_{-0.06}$~\cite{CMS:2020gsy}) and  ATLAS ($\mu =1.06 \pm 0.07$~\cite{ATLAS:2020qdt}) 
one finds the present bounds are already close to the HL-LHC expected ones:
\beq
\mathbf{ATLAS}: 
\quad   \delta y_d < 400,
\quad   \delta y_u < 820,  
\quad   \delta y_s < 19, 
\nonumber
\eeq
\beq
\label{eq:mu-now}
\mathbf{CMS}: 
\quad   \delta y_d < 450, 
\quad   \delta y_u < 930, 
\quad   \delta y_s < 22,  
\eeq
thanks to the small measured upward fluctuation of the Higgs signal strength in ATLAS. 
We consider very encouraging the fact that the sensitivity of the triboson analysis to the first generation Yukawas, cf.~Table~\ref{tab:summary}, is comparable to that in Eq.~\eqref{eq:mu-future}. 
Moreover, many new physics effects may affect the total Higgs signal strength, and cancellations between them are possible.
Including the triboson input in the global Higgs fits will allow to lift degeneracies in the parameter space and obtain more robust constraints on the Higgs couplings. 
We also note that it will be challenging  to significantly improve the sensitivity displayed in Eq.~\eqref{eq:mu-future} at the LHC or future hadron colliders, due to QCD and PDF uncertainties affecting the SM theoretical prediction of the Higgs signal strength.  
This is in contrast with the sensitivity of the triboson analysis, 
which can be significantly improved by upgrading to FCC-hh.  
On the other hand, the sensitivity of the HL-LHC triboson analyses to the strange Yukawa is an ${\cal O}(10)$ factor weaker than the one in Eq.~\eqref{eq:mu-future}. 
Other theoretical ideas (or the FCC-hh) are needed to have a realistic chance of observing effects of enhanced $y_s$.   

We may also compare the triboson sensitivity to $\delta y_u$ and $\delta y_d$ to those projected in other theoretical analyses in the literature, besides that of Higgs signal strength measurements. 
Enhanced light quark Yukawa couplings lead to a distortion of the $p_T$ and rapidity distributions of the $p p \to h$ cross section with respect to the SM predictions due to a larger relative contribution of the $q \bar q \to h$ and $q g \to q h$ processes. Assuming the SM predictions can be controlled at the required level of accuracy, this can be explored at the HL-LHC to set the bounds $\delta y_d \lesssim 380$, $\delta y_u \lesssim 640$~\cite{Soreq:2016rae} (see also~\cite{Bonner:2016sdg,Cohen:2017rsk}), which are comparable to the triboson ones, yet depend on a different set of assumptions.
In the SMEFT, the operators in~Eq.~\eqref{eq:TH_d6} that lead to modified Higgs Yukawa couplings also generate analogous vertices with two (and three) Higgs bosons.  
Therefore one can constrain Yukawa couplings via Higgs pair production, leading to  
$\delta y_d \lesssim 850$, $\delta y_u \lesssim 1200$~\cite{Alasfar:2019pmn} at the HL-LHC.    
Precision measurements of the charge asymmetry of the $W^\pm h$ associated  production could lead to the HL-LHC bounds $\delta y_d \lesssim 1300$, $\delta y_u \lesssim 2900$~\cite{Yu:2016rvv}. 
A large $y_u$ would also enhance $q\bar{q}$-initiated contributions to the $p p \to h \gamma$ process, 
from which the bound $\delta y_u \lesssim 2100$ could be set at the HL-LHC~\cite{Aguilar-Saavedra:2020rgo}. 
Finally, the exclusive Higgs decays $h \to \rho \gamma$ currently probe $\delta y_{u,d}$ of order $10^6$~\cite{Kagan:2014ila,Aaboud:2017xnb}, far from the rest of proposed light quark Yukawa probes. 
We remark that a synergy of several different probes of light Yukawa couplings will be crucial for lifting the various degeneracies in the vast parameter space of the SMEFT.

\section{Conclusion}
\label{sec:conclusion}

\begin{table}[h]
\centering
\begin{tabular}{|c|c|c|c|c|c|c|}
\hline
& \multicolumn{3}{c|}{$WWW$} & \multicolumn{3}{c|}{$ZZZ$} \\
\hline
 & $\ell^\pm \ell^\pm +2\nu +2j$ & $\ell^\pm \ell^\pm \ell^\mp +3\nu$ & Comb. & $4\ell + 2\nu$ & $4\ell + 2j$ & Comb.\\
\hline
$\left. \delta y_d \right .$ & 430 (36) & 840 (54) & 420 (34) & 1500 (65) & 1300 (93) & 1100 (60) \\
\hline
$\left. \delta y_u\right . $ & 850 (71) & 1700 (110) & 830 (68) & 2300 (100) & 1800 (140) & 1600 (92) \\
\hline
$\left. \delta y_s\right .$ & 150 (13) & 230 (33) & 140 (13) & 300 (12) & 290 (16) & 250 (11) \\
\hline
\end{tabular}
\caption{Summary of the projected 2$\sigma$ sensitivity to $\delta y_q$ at the HL-LHC (FCC-hh) for the sub-channels considered in this study.}
\label{tab:summary}
\end{table}

\begin{figure}[t]
\centering
\includegraphics[width=0.85\textwidth]{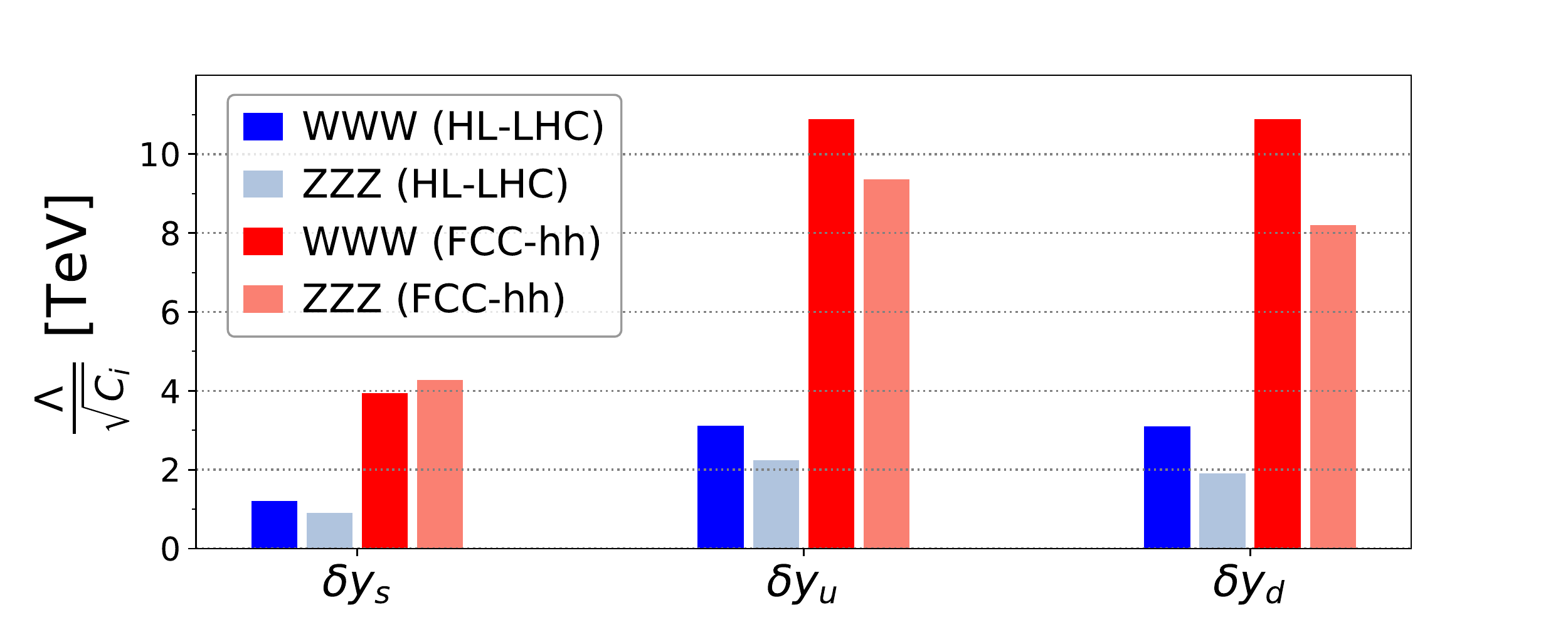} 
\caption{\small Projected 2$\sigma$ reach at the HL-LHC (blue) and FCC-hh (red) on the dimension-6 operator scale $\Lambda$ with Wilson coefficient $C_i$ for the up, down, and strange Yukawa operators in \eqref{eq:TH_d6}. The darker shades are for the combination of the $p p \to WWW \to \ell^\pm \ell^\pm +2\nu +2j$ and $p p \to WWW \to \ell^\pm \ell^\pm \ell^\mp +3\nu$ channels and lighter shades denote the combination of $p p \to ZZZ \to 4l + E_T^\text{miss}$ and $p p \to ZZZ \to 4l + 2j$. }
\label{fig:scaleplot}
\end{figure}

In this study we considered the sensitivity of triboson production to modifications of the up, down and strange Yukawa couplings. Unlike previously suggested probes that rely on on-shell Higgs decays, our proposal makes use of energy growth due to modifications of the off-shell Higgs couplings. We showed that the current CMS triboson analysis constrains $\delta y \lesssim \mathcal{O}(1000)$ at 2$\sigma$ but that this can be improved by an order of magnitude with more targeted cuts to $\delta y \lesssim \mathcal{O}(100)$ at HL-LHC, and, furthermore, to $\delta y \lesssim \mathcal{O}(10)$ at a future 100 TeV collider such as FCC-hh. A summary of the projected bounds are given in Table~\ref{tab:summary}. The corresponding dimension-6 operator scales $\Lambda/\sqrt{C_i}$, defined from~\eqref{eq:TH_d6} as $Y_i = C_i \, v^2/\Lambda^2$ with Wilson coefficient $C_i$, are shown in Fig.~\ref{fig:scaleplot} for the combination of the two $WWW$ sub-channels and the two $ZZZ$ channels. The former are the most sensitive at the HL-LHC, though we note that the $ZZZ$ channel's sensitivity will become comparable at FCC-hh. While these constraints are at the individual level, switching on one operator at a time while setting the others to zero, they can give an indication of the sensitivity of the measurements. In future work, the effects of other SMEFT operators could also be taken into account in a more general analysis, as well as NLO corrections to the differential distributions. 

The experimentally allowed ranges of the light Yukawas are still unconstrained by several orders of magnitude. This can affect other observables (see e.g.~Ref.~\cite{Bishara:2015cha}) and introduce degeneracies in global fits. Moreover, our lack of understanding of the pattern of Higgs couplings motivates probing even unnaturally large enhancements in the light Yukawas that may be difficult to obtain without tuning (see Refs.~\cite{Porto:2007ed,Giudice:2008uua,Bishara:2015cha, Bauer:2015fxa, Bauer:2015kzy} for some examples of specific models). It is therefore important to reduce the experimental uncertainty as much as possible in the future. The study proposed here can be further refined, including more channels and combining related processes, in order to maximise the experimental information available and widen our window onto the mysterious Higgs sector.

\subsection*{Acknowledgments}
We thank Elena Venturini for related collaboration and discussions, the organisers of the Les Houches workshop for the stimulating environment where this project was initiated, Hannsj\"{o}rg Weber for correspondence regarding the CMS analysis and Francesco Riva for correspondence regarding~\cite{Henning:2018kys}. Feynman diagrams were drawn using {\sc TikZ-Feynman}~\cite{Ellis:2016jkw}. KT is supported by the US Department of Energy grant DE-SC0010102. TY is supported by a Branco Weiss Society in Science Fellowship and partially supported by STFC consolidated grant ST/P000681/1. NV is supported by the ERC grant NEO-NAT. AF is partially supported by the Agence Nationale de la Recherche (ANR) under grant ANR-19-CE31-0012 (project MORA). JMN was supported by Ram\'on y Cajal Fellowship contract RYC-2017-22986, and also acknowledges support from the Spanish MINECO's ``Centro de Excelencia Severo Ochoa" Programme under grant SEV-2016-0597, from the European Union's Horizon 2020 research and innovation programme under the Marie Sklodowska-Curie grant agreement 860881 (ITN HIDDeN) and from the Spanish Proyectos de I$+$D de Generaci\'on de Conocimiento via grant PGC2018-096646-A-I00. 

\begin{appendix}

\end{appendix}

\bibliographystyle{JHEP}
\bibliography{biblio.bib}

\end{document}